\documentclass[twocolumn,letterpaper]{IEEEAerospaceCLS}  

\usepackage[]{graphicx}    
\usepackage[colorlinks=false, urlcolor=blue, linkcolor=black, citecolor=black,hidelinks]{hyperref} 
\newcommand{\ignore}[1]{}  
\usepackage[numbers, square]{natbib}
\usepackage{amsmath,amstext,amssymb} 

\normalsize 
\usepackage{booktabs}      
\usepackage{cleveref} 	   
\usepackage{tikz} 
\usepackage{float}
\usepackage{algorithm} 
\usepackage[noend]{algpseudocode} 
\begin{document}
\title{Scalable Ground Station Selection for \\ Large LEO Constellations}

\author{Grace Ra Kim, Duncan Eddy, Vedant Srinivas, and Mykel J. Kochenderfer\\ 
Stanford University\\
496 Lomita Mall \\
Stanford, CA 94305 \\
\{gkim65, deddy, vedants8, mykel\}@stanford.edu
\thanks{\footnotesize 979-8-3315-7360-7/26/$\$31.00$ \copyright2026 IEEE}              
}

\maketitle

\thispagestyle{plain}
\pagestyle{plain}

\maketitle

\thispagestyle{plain}
\pagestyle{plain}

\begin{abstract}
\textbf{Effective ground station selection is critical for low-Earth orbiting (LEO) satellite constellations to minimize operational costs, maximize data downlink volume, and reduce communication gaps between access windows. Traditional ground station selection typically begins by choosing from a fixed set of locations offered by Ground Station-as-a-Service (GSaaS) providers, which helps reduce the problem scope to optimizing locations over existing infrastructure. However, finding a globally optimal solution for stations using existing mixed-integer programming methods quickly becomes intractable at scale, especially when considering multiple providers and large satellite constellations. To address this issue, we introduce a scalable, hierarchical framework that decomposes the global selection problem into single-satellite, short time-window subproblems. Optimal station choices from each subproblem are clustered to identify consistently high-value locations across all decomposed cases. Cluster-level sets are then matched back to the closest GSaaS candidate sites to produce a globally feasible solution. This approach enables scalable coordination while maintaining near-optimal performance. We evaluate our method's performance on synthetic Walker-Star test cases (1–10 satellites, 1–10 stations), achieving solutions within 95\% of the global IP optimum for all test cases. Real-world evaluations on Capella Space (5 satellites), ICEYE (40), and Planet’s Flock (96) show that while exact IP solutions fail to scale, our framework continues to deliver high-quality site selections.}
\end{abstract} 

\tableofcontents
\section{Introduction}

The rapid expansion of commercial Earth observation constellations has increased the demand for ground station infrastructure capable of supporting high-throughput, low-latency, and continuous downlink services. Traditional fixed-site ground networks, originally designed for government and scientific missions, are increasingly strained by the scale and responsiveness required by emerging commercial operators~\cite{cleverly2021evolution,carvalho2019optimizing,vasisht2020distributed}. To address this, Ground-Station-as-a-Service (GSaaS) platforms have emerged as a promising paradigm, offering shared, flexible, and globally distributed access to ground infrastructure. Yet, the design and selection of GSaaS sites to support rapidly growing constellations remain an open challenge. Integer programming (IP) formulations, while globally optmial, exhibit exponential growth in complexity as constellation sizes, ground station networks, and scheduling horizons increase, resulting in prohibitive solve times and poor scalability beyond small problems. In particular, optimizing objectives such as minimizing communication gaps across diverse orbits and providers quickly becomes computationally intractable~\cite{eddy_optimal_2024}, motivating the need for scalable approximation methods tailored to GSaaS site selection.

Recent work on ground station placement and site selection has primarily focused on optimization-based approaches. Eddy et al.~\cite{eddy_optimal_2024} present a comprehensive IP formulation for optimal GSaaS selection, with objectives minimizing mission costs while respecting constraints such as for minimum data downlink capacity and maximum communication gaps. Kopacz et al.~\cite{kopacz2019optimized} explore the use of genetic algorithms, developing a multi-objective approach that accounts for infrastructure availability and constellation connectivity, determining optimal station placement for mega-constellations using revenue-based fitness functions. Evolutionary algorithms have further been used to optimize the number and deployment of ground tracking stations for low Earth orbit (LEO) constellations, maximizing observation coverage and minimizing satellite position dilution of precision~\cite{Kralfallah_Optimizing_2024}. Despite these advances, scaling these solutions to support large, rapidly growing satellite constellations remains a significant challenge, as both exact and heuristic methods encounter computational limitations when faced with large networks, multiple satellites, and extended scheduling horizons.


Outside of the space domain, optimization approaches for location selection and resource scheduling generally fall into exact, heuristic, and metaheuristic categories. Exact formulations such as mixed-integer linear programming (MILP) and IP offer precise control over capacity, connectivity, and cost constraints, but quickly become impractical as problem sizes and combinatorial possibilities grow~\cite{owen1998strategic}. Heuristic and metaheuristic methods, including genetic algorithms and particle swarm optimization, scale more effectively by iteratively searching large solution spaces, yet still face limits in extreme cases~\cite{Azeem_sub_2019, Wei_multi_2022}. Related techniques, such as stochastic programming, swarm intelligence, and nonlinear optimization, have been applied in domains like sensor deployment, ambulance routing, and multi-drone coordination~\cite{cardei2006energy, tlili2017swarm, hu_joint_2018}. Across these areas, scalability remains a core bottleneck: solution quality and runtime degrade rapidly as instance sizes increase, underscoring the challenge of applying such methods to ground network design.

One common strategy for managing this scalability is to reduce problem complexity before optimization. Clustering techniques such as $k$-means, have often been applied in facility location and related optimization problems, primarily as a pre-processing step to reduce input complexity~\cite{franti2025designing,imai2023analysis}. By grouping candidate sites or demand points, clustering simplifies large-scale formulations and makes them more tractable for exact or heuristic solvers. While effective for input reduction, most applications stop at initialization. Only recently has clustering been explored downstream, to aggregate or reallocate optimized assignments for improved scalability or interpretability~\cite{buijs2025effective}. 
However, post-solve clustering remains rare, particularly in the context of reducing complexity in large-scale problems for GSaaS site selection.

In this work, we overcome the computational challenges faced by traditional IP solvers on large GSaaS site selection problems by applying a scalable decomposition, post-solve clustering, and matching framework. Our method first decomposes the full GSaaS IP formulation into smaller subproblems with shorter overlapping time windows, single-satellite problems, and reduced GSaaS site candidate locations. The resulting solutions are aggregated via a clustering-based pipeline employing either DBSCAN~\cite{DBSCAN} and $k$-Medoids~\cite{kaufman1990partitioning}, where for DBSCAN, clusters are matched to the closest ground station networks using Hungarian assignments to construct feasible, near-optimal global placements. This approach effectively combines the solution quality of integer programming with the scalability benefits of clustering, enabling efficient exploration of extensive candidate site sets that typically overwhelm monolithic IP formulations.

We validate our method through two studies. First, we assess optimality on smaller scenarios using a synthetic Walker-Star constellation, selecting sites from a subset of GSaaS providers (KSAT and Atlas Space). For the maximum data downlink objective, our DBSCAN clustering with Hungarian matching approach achieves solutions within 99.3\% of the global IP solver optimum across all scenarios. For the minimum maximum-gap objective, over 94\% of problems reach at least 95\% of the true optimal value. After establishing performance on small-scale scenarios, we evaluate higher-complexity problems on real commercial Earth observation constellations, including Capella Space, ICEYE, and Planet Labs' Flock spacecraft. Again, for maximum data downlink, all scenarios achieve near-optimal results within 97\% of the IP solution. While the full IP models become intractable for the minimum maximum-gap objective in larger constellations, our method still produces high-quality solutions with minimal degradation. For our largest test case with Planet Labs' 96-satellite Flock constellation, we find that the maximum gap increases by no more than 0.31 hours from the per-satellite decomposed solution. By combining IP decomposition, clustering, and matching, we provide a practical path toward scalable GSaaS site selection.
\section{Problem Formulation}
\label{sec:problem_formulation}

We define the GSaaS site selection problem as selecting $n$ stations from candidate locations to optimize a single mission performance metric for a given satellite constellation. The stations must satisfy operational constraints, ensuring feasible, reliable, and efficient mission operations. Because the problem involves binary decisions over station selection and contact opportunities computed between site locations and spacecraft, it is naturally expressed as an integer program. While this IP formulation precisely captures the problem and can be solved for small instances, the combinatorial growth of variables and constraints makes direct optimization intractable for many networks. In the following section, we present the full IP formulation, which serves as the basis for the scalable decomposition, clustering, and matching framework introduced in \Cref{sec:scalable}.

\subsection{IP Formulation}

Building on the formulation presented by Eddy et al.~\cite{eddy_optimal_2024}, our ground station optimization problem considers a set of candidate providers $\mathcal{P}$, where each provider is represented as a tuple $(p, P)$. Here, $p \in \{0,1\}$ is a binary decision variable indicating whether the provider is selected $p=1$ or not $p=0$, and $P$ is used to denote the provider and variables related to the provider. Each provider offers a set of ground station locations, denoted ${\mathcal{L}}^P$. Each location inside $\mathcal{L^P}$ is represented as a tuple $(l, L)$, where $l \in \{0,1\}$ is a binary inclusion variable and $L$ contains the location’s fixed characteristics, including its data rate $L_{dr}$. We allow the final network to include stations from multiple providers. The full set of possible locations from all providers is denoted as $\mathcal{L}$. 

Network selection is performed with respect to the satellite constellation $\mathcal{S}$, where each satellite $S\in\mathcal{S}$ has a fixed data rate $S_{dr}$. From $\mathcal{P}$ and $\mathcal{S}$, we can compute the set of all possible satellite-to-ground contacts $\mathcal{C}$. Network performance depends not only on which stations are included but also on which satellite-to-ground contacts are scheduled. We evaluate the optimality of each candidate ground network by assessing the quality of its available contacts and selecting the subset that maximizes the performance of specific mission-level objectives, such as total downlinked data. Subsets of contacts associated with a particular provider, station, or satellite are denoted $\mathcal{C}^P \subset \mathcal{C}$, $\mathcal{C}^L \subset \mathcal{C}$, and $\mathcal{C}^S \subset \mathcal{C}$, respectively. Each contact $(c, C) \in \mathcal{C}$ has a binary decision variable $c \in \{0,1\}$ indicating whether it is scheduled, as well as fixed attributes including start time $C^{start}$, end time $C^{end}$, and duration $C_{duration} = C^{end} - C^{start}$. Contact data rate 
\begin{equation}
\begin{aligned}
	C_{dr} = \min(L_{dr},S_{dr})
\end{aligned}
\label{eqn:contact_datarate}
\end{equation}
is the minimum of the station and satellite data rates. By jointly optimizing station and contact selection, the integer program ensures that the resulting network configuration meets mission objectives and operational constraints.

In principle, the full contact set $\mathcal{C}$ should be computed over the entire mission horizon to determine the optimal set of ground station network locations and contacts to schedule for the mission. This mission horizon is defined by the start and end times $t_{opt}^{start}$ and $t_{opt}^{end}$, with total duration $T_{opt} = t_{opt}^{end} - t_{opt}^{start}$. In practice, however, long-term trajectory forecasts for LEO satellites are unreliable due to orbital perturbations. To overcome this limitation, we adopt a surrogate optimization approach inspired by Eddy et al.~\cite{eddy_optimal_2024}, in which ground station placement is optimized over shorter, representative simulation intervals—typically 7 to 10 days—that capture multiple orbital periods and yield a statistically meaningful approximation of contact opportunities across the mission. The surrogate interval is defined by the simulation start and end times $t_{sim}^{start}$ and $t_{sim}^{end}$, with duration $T_{sim} = t_{sim}^{end} - t_{sim}^{start}$.

\subsection{Objective Functions}

Having defined the decision variables and relevant problem parameters, we now formulate the objective function that guides the optimization of our GSaaS site selection. In this work, we consider two main objectives: the maximum data downlink objective, which seeks to maximize the total amount of data transferred from satellites to the ground network, and the minimum maximum-gap objective, which aims to minimize the longest downtime between consecutive contacts for any satellite.

The \textit{maximum data downlink objective} seeks to design a ground station network that maximizes the total volume of data transmitted from the satellite constellation over the mission. This objective is particularly relevant for missions where data return is the primary priority and communication opportunities are limited. To compute this objective, we define the total data downlinked over the entire mission as
\begin{equation}
\begin{aligned}
	\frac{T_{opt}}{T_{sim}}\sum_{P \in {\mathcal{P}}}\sum_{L \in {\mathcal{L}}^P}\sum_{(c,C) \in {\mathcal{C}}^L}C_{dr}C_{duration}c
\end{aligned}
\label{eqn:obj_max_data}
\end{equation}
where each contact contributes its data rate $C_{dr}$ multiplied by its duration $C_{duration}$ if it is scheduled. That is, if $c = 1$. The factor $T_{opt}/T_{sim}$ scales the objective from the surrogate simulation interval to the full mission duration, providing an estimate of the total data volume downlinked over the mission.

The \textit{minimum maximum-gap objective} seeks to minimize the longest time interval between consecutive contacts for each satellite, including the gaps at the beginning and end of the mission window. This objective is particularly relevant for operationally responsive missions, such as Earth observation, where minimizing latency in downlinking data or uploading new tasking instructions is critical. By reducing the maximum gap between contacts, the network ensures that all satellites maintain frequent and reliable ground communication.

To model this objective, we introduce an auxiliary variable, $G_{\max}$, representing the maximum contact gap across all satellites. For each satellite $S \in \mathcal{S}$, we define binary variables $y_{ij}$ indicating that contact $j$ immediately follows contact $i$ in the schedule, as well as auxiliary binary variables $y_{i,first}$ and $y_{i,last}$ to mark the first and last contacts of the satellite relative to the start and end of the mission window. The full optimization objective is
\begin{subequations}
\begin{align}
    &\underset{\mathbf{x}}{\text{minimize}} && G_{\max} \label{eqn:obj_min_max_gap_1} \\
    &\text{s.t.} && \sum_{j > i} y_{ij} + y_{i,last} = c_i, \quad \forall (c,C) \in \mathcal{C}^S, S\in \mathcal{S} \label{eqn:only_one_next}\\
    & && \hspace{3em} i,j \in \{0,\ldots,\lvert\mathcal{C}^S\rvert\} \;\text{s.t. } \label{eqn:orderContacts}\\
    & && C^{start}_j > C^{end}_i \label{eqn:contactRule} \\
    & && (C^{start}_j - C^{end}_i)\, y_{ij} \le G_{\max} \label{eqn:gap_between_contacts} \\
    & && (C^{start}_i - t_{sim}^{start})\, y_{i,first} \le G_{\max} \label{eqn:first_contact_gap} \\
    & && (t_{sim}^{end} - C^{end}_i)\, y_{i,ast} \le G_{\max} \label{eqn:last_contact_gap} \\
    & && y_{ij} \le c_i, \quad y_{ij} \le c_j \label{eqn:link_contacts} \\
    & && y_{i,first} \le c_i, \quad y_{i,last} \le c_i \label{eqn:link_first_last} \\
    & && y_{ij},\, y_{i,first},\, y_{i,last},\, c_i \in \{0,1\} \label{eqn:binary_variables} \\
    & && \sum_{i \in \mathcal{C}^S} c_i \ge 1 \label{eqn:at_least_one_contact} \\
    & && \sum_{i \in \mathcal{C}^S} y_{i,first} = 1, \quad \sum_{i \in \mathcal{C}^S} y_{i,last} = 1 \label{eqn:exactly_one_first_last}
\end{align}
\label{eqn:obj_min_max_gap_full}
\end{subequations}

All gaps are determined from sequentially scheduled contacts. Constraint \eqref{eqn:only_one_next} enforces that each scheduled contact $c_i$ has exactly one successor contact $c_j$, unless it is designated as the last contact. This is handled by the binary variable $y_{i,last}$ which allows a contact to terminate the sequence. Constraint \eqref{eqn:orderContacts} and \eqref{eqn:contactRule} are included so that contacts are sorted in time order. Constraint \eqref{eqn:gap_between_contacts} ensures that $G_{\max}$ is at least as large as the gap between any two consecutive contacts. Constraints \eqref{eqn:first_contact_gap} and \eqref{eqn:last_contact_gap} similarly bound the initial and final gaps between the simulation window and the first and last selected contacts by $G_{\max}$. Constraint \eqref{eqn:link_contacts} links each chosen transition $y_{ij}$ to the corresponding contact selection variables $c_i$ and $c_j$, while Constraint \eqref{eqn:link_first_last} ensures consistency by tying the first and last decision variables to the correct first selected and last selected contacts. Constraint \eqref{eqn:at_least_one_contact} requires at least one contact to be chosen for each satellite, and Constraint \eqref{eqn:exactly_one_first_last} ensures that exactly one first contact and one last contact are selected. 

\subsection{Constraint Functions}

In addition to the objective functions, a set of constraints must be introduced to ensure that the optimization problem produces feasible and meaningful solutions. These constraints fall into two categories: (1) structural constraints that link contacts, locations, and providers to construct feasible solutions and (2) systems-engineering constraints that reflect practical mission requirements.

First, we impose structural constraints to link the selection of contacts to their corresponding ground stations and providers. If a contact $c$ is scheduled, then the associated ground station location variable $l$ must also be active:
\begin{equation}
\begin{aligned}
   	\sum_{(c, C) \in {\mathcal{C}}^L}c \leq \lvert\mathcal{C}^L\rvert l, \; \forall \; (l,L) \in {\mathcal{L}}
 \end{aligned}
\label{eqn:constraint_location_selection}
\end{equation}
Similarly, if any location $l$ belonging to a provider $p$ is selected, the provider itself must be included in the solution
\begin{equation}
\begin{aligned}
   	\sum_{(l, L) \in {\mathcal{L}}^P}l  \leq \lvert\mathcal{L}^P\rvert p, \; \forall \; (p, P) \in {\mathcal{P}}
 \end{aligned}
\label{eqn:constraint_provider_selection}
\end{equation}
In addition to these structural constraints, we introduce three systems-engineering constraints to capture realistic mission requirements. 

The first is the \textit{satellite contact exclusion constraint}, which prevents a satellite from being scheduled to communicate with more than one ground station at the same time. This constraint should be applied unless the spacecraft can support simultaneous downlinks. The constraint can be expressed as
\begin{equation}
\begin{aligned}
	& c_i + c_j \leq 1 \; \\
   	& \hspace{2em} \forall \; S \in {\mathcal{S}}, i,j \in \{0,\ldots,|{\mathcal{C}}^S|\}, j > i \; \text{s.t.} \\
   	& C_{start,i} \leq C_{end,j} \\
   	& C_{start,j} \leq C_{end,i} \\
\end{aligned}
\label{eqn:constraint_contact_exclusion}
\end{equation}

The second is \textit{minimum contact duration constraint}, where short-duration contacts that may not be feasible for use are filtered out from selection
\begin{equation}
\begin{aligned}
   	& x = 0 \; {if} \; C_{end} - C_{start} < t_{min}, \forall \; (c,C) \in {\mathcal{C}}
\end{aligned}
\label{eqn:constraint_min_duration}
\end{equation}
where any contact with a duration shorter than a threshold $t_{min}$ is excluded from scheduling.

Finally, the \textit{station number constraint} enforces bounds on the number of ground stations selected in the final design, ensuring that the solution respects architectural limits
\begin{equation}
\begin{aligned}
   	& \sum_{l \in {\mathcal{L}}} l \geq M_{min} \\
   	& \sum_{l \in {\mathcal{L}}} l \leq M_{max} \\
\end{aligned}
\label{eqn:constraint_required^location}
\end{equation}
ensuring that at least $M_{min}$ stations are selected and no more than $M_{max}$ are selected. As we are selecting a network of size $n$, $M_{min} = M_{max} =n$. Together, these constraints ensure both logical consistency (e.g., linking contacts to locations and providers) and mission feasibility (e.g., preventing overlapping satellite contacts, eliminating unusable opportunities, and bounding network size).

This IP formulation exactly captures the GSaaS site selection problem, and for small instances, it can be solved using established optimization solvers such as Gurobi~\cite{gurobi} and COIN-OR~\cite{lougee2003common}, which provide certificates of optimality or infeasibility. However, for realistic constellation and network scales, the total number of binary decision variables and constraints is dominated by the contact sequencing terms, scaling approximately as $O(\lvert \mathcal{S}\rvert  \cdot \lvert \mathcal{C}^S\rvert ^2)$, where $\lvert \mathcal{S}\rvert$ is the number of satellites and $\lvert \mathcal{C}^S\rvert$ is the number of contacts per satellite. Additional terms scale linearly with the number of candidate locations and providers, making the full IP quickly intractable for large constellations or dense candidate networks. To address this challenge, we present a decomposition, clustering, and matching framework that preserves the fidelity of the IP in smaller subproblems while enabling scalable optimization.
\begin{figure*}[tbp]
    \centering
    \includegraphics[width=0.95\linewidth]{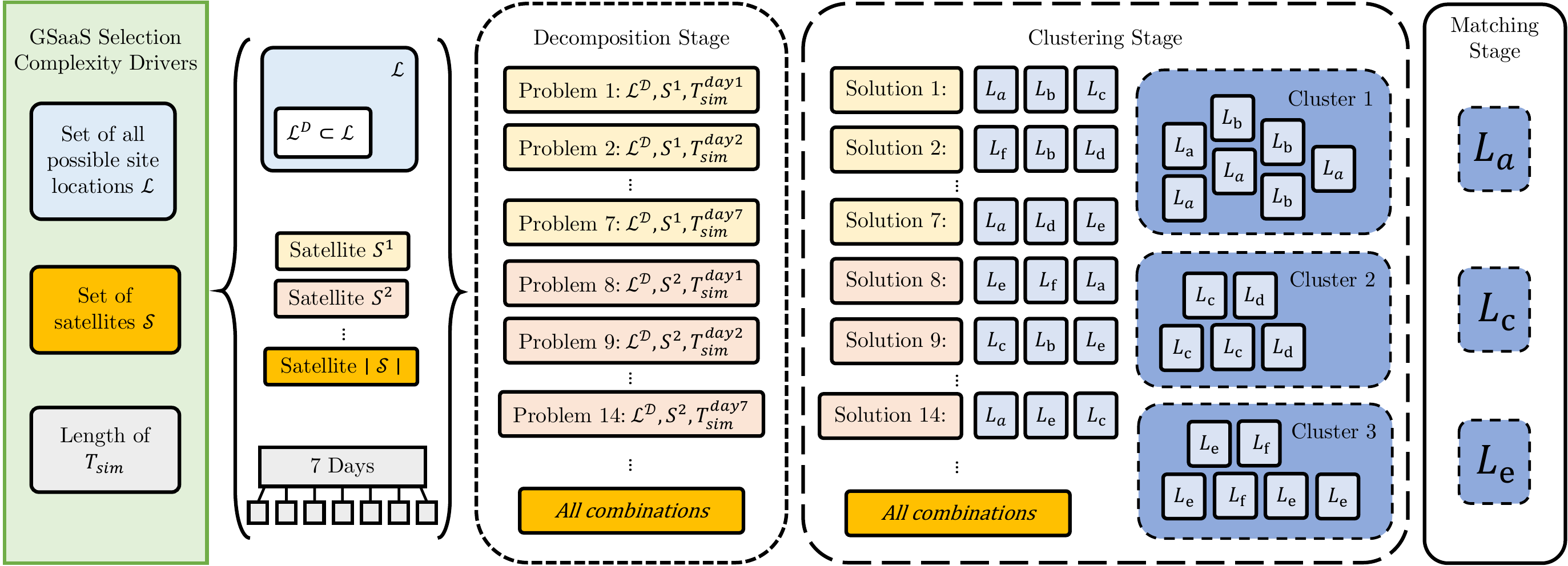}
    \caption{Scalable GSaaS Placement via Decomposition and Clustering}
    \label{fig:clusteringMech}
    \vspace{-2em}
\end{figure*}

\section{Scalable GSaaS Placement}
\label{sec:scalable}

Solving the full GSaaS site selection IP formulation for large constellation and network scales results in severe scalability challenges. In particular, the complexity grows along three fronts: (i) the number of candidate ground stations, (ii) the size of the satellite constellation, and (iii) the number of satellite-to-ground station contact opportunities, which grows with longer lengths of simulated time horizons $T_{\text{sim}}$. Each potential contact introduces scheduling variables with corresponding constraints on coverage, timing, and resource allocation, leading to a combinatorial explosion in the number of constraints and decision variables. When solved as a single monolithic IP, this quickly becomes intractable, resulting in excessive runtime and memory demands that prevent the formulation from being applied to realistic constellations and dense ground station networks.

To overcome these scaling barriers, we develop a decomposition and clustering framework for GSaaS placement. The approach proceeds in three stages, illustrated in \Cref{fig:clusteringMech}. First, the original IP is decomposed into smaller, tractable subproblems by partitioning along the main drivers of complexity: candidate ground stations, constellation size, and contact opportunities. Second, the site selections from these subproblems are clustered to extract representative ground station sites that approximate those of the full IP formulation. Finally, these representative selections are matched to actual GSaaS locations, producing globally feasible placements.

\subsection{Decompositions to Reduce Problem Complexity}

The first stage of our framework decomposes the initial GSaaS IP formulation into smaller, tractable subproblems. The motivation is to isolate the main drivers of complexity: candidate ground stations, constellation size, and contact opportunities, so that each subproblem remains solvable within reasonable time and memory limits.

\subsubsection{\textbf{Candidate GSaaS Pool Restriction and Expansion}}
The full candidate set of GSaaS sites $\mathcal{L}$ can be very large, creating a correspondingly large IP with high variable and constraint counts. To reduce complexity, we first restrict the problem to a smaller subset $\mathcal{L}^D \subset \mathcal{L}$ consisting of geographically diverse locations. This restriction yields faster solve times while preserving representative geographic coverage. After solving the restricted subproblems, the results are expanded back to the full set $\mathcal{L}$ during the matching phase to select final sites over the complete candidate pool.

\subsubsection{\textbf{Temporal Decomposition via Overlapping Windows}}
We also mitigate computational complexity by decomposing the problem along the temporal dimension. Over a long simulation horizon $T_{\text{sim}}$, the number of satellite-to-ground station contacts can become extremely large, resulting in an IP with excessive scheduling variables and constraints. To address this, we partition the simulation horizon $T_{sim}$ into a set of overlapping time windows $\mathcal{T} = \{T_{sim}^1, T_{sim}^2, \dots, T_{sim}^W\}$, each of length $\Delta t$. Each window defines a smaller subproblem, reducing the number of contacts and scheduling variables considered per IP instance and enabling faster solves. Overlapping windows ensure continuity of the solution across boundaries, preventing fragmentation of contact schedules and maintaining feasibility. In our experiments, we set $\Delta t$ to one day, with a 12-hour overlap between consecutive windows, resulting in $W=2T_{sim}/\Delta t-1$ subproblems. The window length and overlap can be adjusted to balance computational efficiency and solution fidelity. In practice, we found that a minimum $\Delta t$ of one day was needed to provide a sufficient number of contacts for effective optimization while maintaining tractable subproblem sizes.

\subsubsection{\textbf{Per-Satellite Decomposition}}
For problems involving large constellations, a final decomposition layer can be added to partition the GSaaS selection problem spatially by satellite. At this stage, each satellite’s contact schedule is treated as an independent subproblem within each time window. In cases where it is advantageous, groups of satellites $\mathcal{S}^G \subset \mathcal{S}$ can also be solved together to balance efficiency and coordination. Decomposing the problem by reducing the constellation size leverages the operational independence of satellites’ communication schedules, substantially reducing the size of each IP and memory requirements. We ensure key structural constraints are enforced within each satellite subproblem, so the resulting schedules remain feasible with optimal selections at the individual satellite level.

\subsection{Clustering-Based Aggregation of Subproblem Solutions}
Once the decomposed subproblems are solved, we aggregate the resulting candidate station selections across every scenario. Each subproblem is defined over the restricted candidate set $\mathcal{L}^D$, a satellite $S$ or satellite group $\mathcal{S}^G$, and a time window $T_{\text{sim}}^j$, for all $S \in \mathcal{S}$ or $\mathcal{S}^D \subset \mathcal{S}$ and $T_{\text{sim}}^j \in \mathcal{T}$. Collecting the solutions produces a dataset of selected locations, with further decompositions of constellations or time windows providing additional data for clustering.

While any clustering method can be applied, we implement DBSCAN \cite{DBSCAN} and $k$-Medoids \cite{kaufman1990partitioning} using Python's \textsc{scikit-learn} \cite{scikit-learn} and \textsc{kmediods}~\cite{schubert2022fast} libraries. DBSCAN groups locations based on spatial density, revealing dominant geographic and network patterns in the placement assignments and naturally identifying dense regions without pre-specifying the number of clusters. The resulting clusters allow us to compute centroids, which are then matched to the closest actual GSaaS sites. In contrast, $k$-Medoids selects actual data points as cluster centers, directly producing representative ground station sites and eliminating the need for a separate matching stage. Using both methods allows us to compare the effectiveness of aggregation with and without the additional matching step.

\subsection{Matching Clusters-to-GSaaS Sites}

Building on the clustering results, we now assign final placements of these cluster centers to actual GSaaS sites across five current GSaaS providers—Atlas Space Operations, AWS Ground Station, KSAT, Leaf Space, and Viasat. We reference the full GSaaS site listing from Eddy et al.~\cite{eddy_optimal_2024}. The matching process uses a Hungarian assignment algorithm~\cite{kuhn1955hungarian} implemented via Python's \textsc{scipy.optimize} library~\cite{2020SciPy-NMeth}, which finds the optimal assignment minimizing total geodesic distance across all pairs. This ensures that the clustered solutions are translated into feasible, existing locations. We also compare matching to the initial candidate pool $\mathcal{L}^D$ versus the expanded full list of sites $\mathcal{L}$. Comparative results of these methods are presented in the experimental section. For completeness, a brief description of the clustering and matching algorithms is provided in the Appendix.
\section{Experiments}

We perform several evaluations of our scalable GSaaS placement framework, focusing on two main areas. First, we compare scalability and accuracy of each stage in our pipeline to the exact integer programming formulations when one-to-one comparisons can be made. These are for small scenarios when the IP solver can converge to the globally optimal solution. Second, we dive into the performance of our scalable decomposition, clustering, and matching solution in realistic scenarios with higher complexity. The complexity arises from both considering a larger number of candidate solutions $\mathcal{L}$ and the size of the satellite constellation $\mathcal{S}$ supported. These experiments are performed on the Capella Space, ICEYE, and Planet Labs' Flock constellations. In these scenarios, we demonstrate the effectiveness and scalability of our decomposition, clustering, and matching framework when the IP formulation alone fails.

Spacecraft dynamics are propagated using the SGP4 propagator \cite{vallado2006revisiting}. The simulation horizon $T_{sim}$ is set to 7 days, split into 1-day windows with 12-hour overlaps, resulting in 14 windows. The full optimization horizon $T_{opt}$ is set to 365 days for all cases. A $10^\circ$ minimum elevation mask is applied to all contact opportunities. Simulations were performed on a workstation with a 64-core 2.0 GHz AMD EPYC 7713 processor. Detailed design parameters for the main simulations are provided in \Cref{tab:design_parameters}.

\begin{table}[htbp]
\centering
\caption{Design parameters used for simulations.} 
\vspace{-1em}
\begin{tabular}{@{}lr@{}}
\toprule
Parameter & Value \\
\midrule
$t^{start}_{sim}$ & 2025-08-22 00:00:00 UTC \\
$t^{end}_{sim}$ & 2025-08-29 00:00:00 UTC \\
$t^{start}_{opt}$ & 2025-08-22 00:00:00 UTC \\
$t^{end}_{opt}$ & 2026-08-22 00:00:00 UTC \\
$C_{dr}$ & 1.2 Gbps \\
$\Delta t$ & 1 day \\
$T_{overlap}$ & 12 hours \\
$t_{\min}$ & 180 seconds \\
\bottomrule
\end{tabular}
\label{tab:design_parameters}

\end{table}

In our experiments, we focus on two primary objectives: maximizing data downlink and minimizing each satellite's maximum contact gap. These objectives capture the trade-offs inherent in designing ground station segments for satellite constellations, and we use the full IP objective function formulation and constraints as described in \Cref{sec:problem_formulation}. Each IP subproblem is solved using Gurobi v11.0.3, and the resulting solutions are fed into our clustering and matching pipeline using the appropriate python \textsc{scikit-learn} and \textsc{scipy.optimize} libraries for final GSaaS site selection.

\subsection{Method Accuracy for Smaller Problem Sizes}

We first evaluate our decomposition framework on small problem instances where an IP solver can obtain exact solutions. For each test case, we generate new problem scenarios by varying the satellite constellation size $\lvert \mathcal{S} \rvert $ and the number of selected ground stations $n$. To do this, we construct a synthetic Walker-Star constellation with the parameters in \Cref{tab:walker_star} to systematically vary constellation size and geometry. The number of orbital planes is varied from 1 to 10, producing constellations ranging from a single satellite to 10 satellites. We further increase problem complexity by varying the number of ground stations $n$ selected from our design candidate list $\mathcal{L}^D\subset\mathcal{L}$, which is restricted to the KSAT and Atlas Space networks. Optimizations with larger candidate site sets $\mathcal{L}$, and problems needing to select larger number of ground stations $n$ yield harder problems. 

\begin{table}[tbp]
\centering
\caption{Walker-Star constellation parameters.} 
\vspace{-1em}
\begin{tabular}{@{}lcr@{}}
\toprule
Parameter & Value & Unit \\
\midrule
Altitude & 781 & km \\
Eccentricity & 0.001 & -- \\
Inclination & 86.4 & degrees \\
\# of Planes & $\{1,2,\dots,10\}$ & -- \\
\# of Sats per Plane & 1 & -- \\
\bottomrule
\end{tabular}
\label{tab:walker_star}
\vspace{-1em}
\end{table}

\begin{figure}[htbp]
  \centering

    \includegraphics[trim={1cm 1cm 0cm 1cm},scale = 0.29]{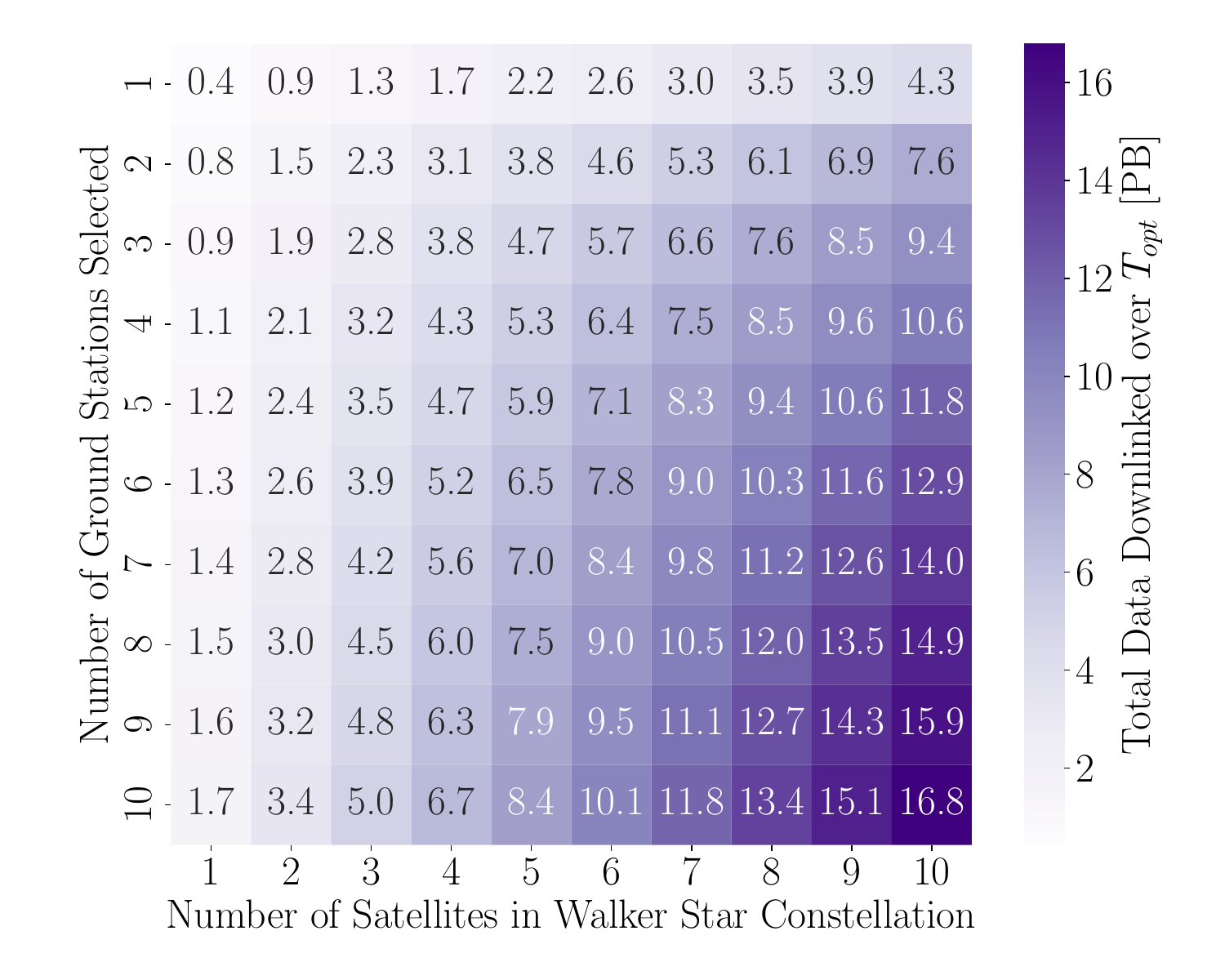}
    \caption{Optimal full-IP solution for maximum data downlink, varying $\#$ of ground stations and satellites in constellation. Benchmark “true” solution, with all original constraints enforced.}
    \label{fig:IP_solmaxdata}
    \vspace{-1em}
\end{figure}

\begin{figure}[htbp]
  \centering

    \includegraphics[trim={1cm 1cm 0cm 1cm},scale = 0.29]{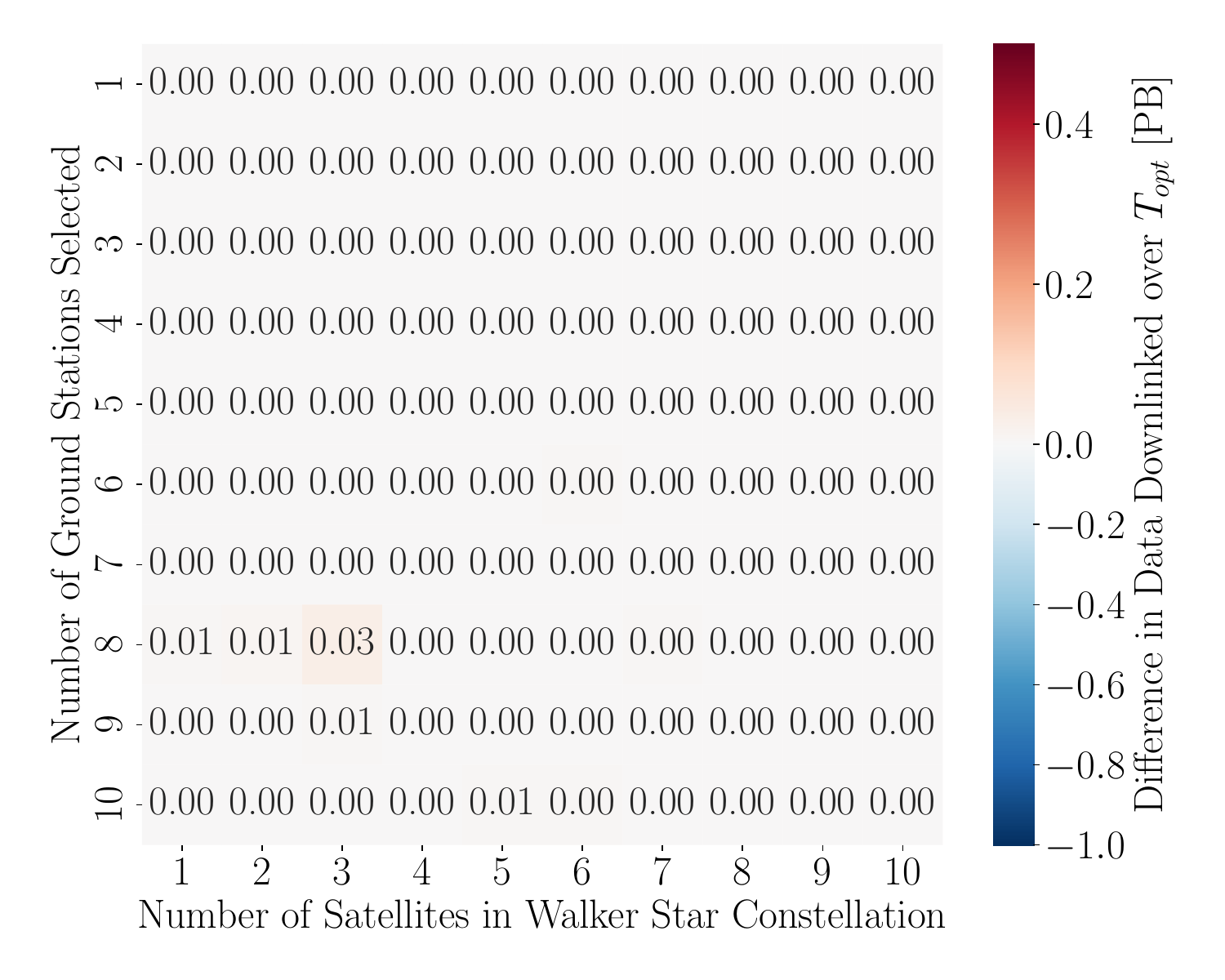}
    \caption{Performance deviation of DBSCAN with Hungarian matching from optimal in \Cref{fig:IP_solmaxdata}, maximum data downlink objective. Decomposition performed for shorter time windows and per-satelite subproblems. All solutions within $99\%$ of optimal.}
    \label{fig:DBSCAN_solmaxdata_final}
    \vspace{-2em}
\end{figure}

Starting with the maximum data objective, the full IP solution in \Cref{fig:IP_solmaxdata} provides the ground-truth optimal total downlink over $T_{opt}$ in petabytes (PB). In \Cref{fig:DBSCAN_solmaxdata_final}, we plot the difference between this optimal solution and the best values achieved by our final DBSCAN with Hungarian matching solution across each test case. This difference heatmap allows us to directly visualize where our methods match, improve upon, or fall short of the IP solution. Since we take the difference from the IP solution, negative values (blue) indicate performance matching or exceeding optimal, while positive values (red) show suboptimality.

\begin{figure}[tbp]
  \centering

    \includegraphics[trim={1cm 1cm 0cm 1cm},scale = 0.29]{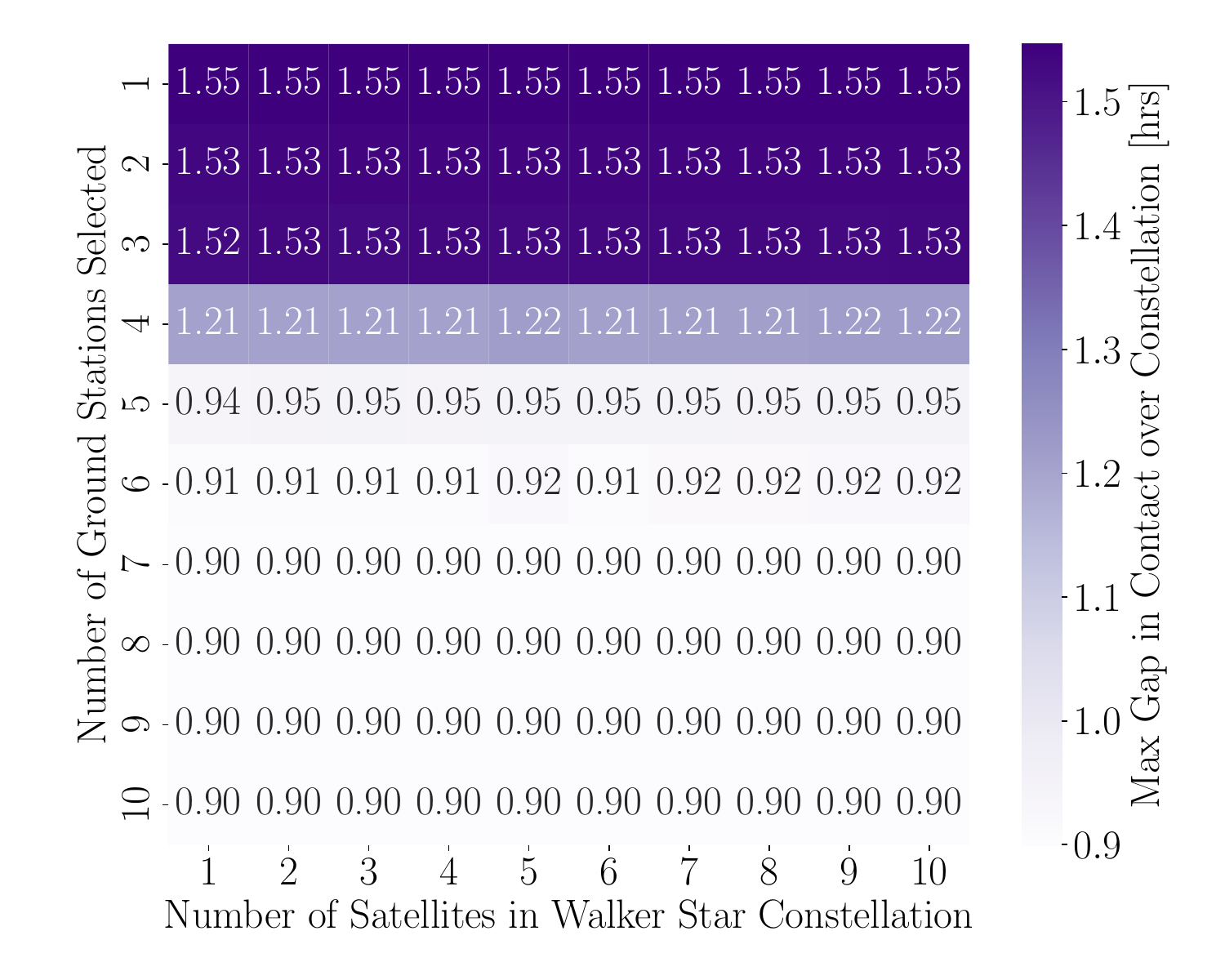}
    
    \caption{Optimal full IP solution for minimum maximum-gap for all satellites as a function of selected $\#$ of ground stations and satellites in constellation. Benchmark “true” solution, with all original constraints enforced.}
    \label{fig:IP_solminmaxgap}
    \vspace{-1em}
\end{figure}

\begin{figure}[tbp]
  \centering

    \includegraphics[trim={1cm 1cm 0cm 1cm},scale = 0.29]{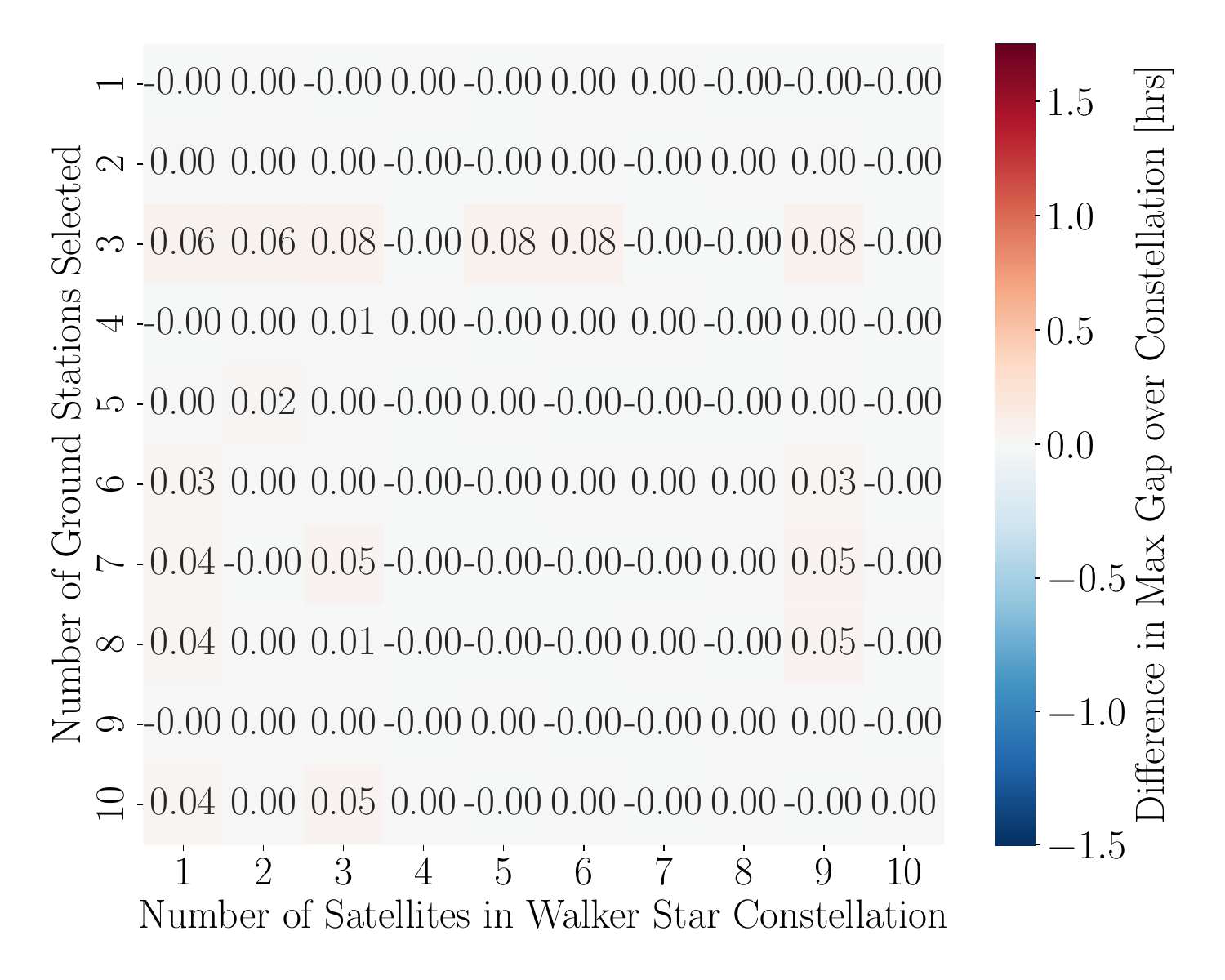}
    \caption{Deviation in performance of DBSCAN and Hungarian matching from optimal full-IP for minimium maximum-gap as a function of selected $\#$ of ground stations and satellites in constellation. Scalable decomposition applied along only time dimensions, with all solutions within $95\%$ of the optimal.}
    \label{fig:DBSCAN_solminmaxgap_final}
    \vspace{-2em}
\end{figure}

We similarly plot the ground-truth IP solution for the minimum maximum-gap objective in \Cref{fig:IP_solminmaxgap}, which displays maximum contact gap over all satellites in hours during over $T_{opt}$. \Cref{fig:DBSCAN_solminmaxgap_final} displays the difference between the optimal solution and our method. 

Comparing the optimal IP solutions against our scalable decomposition, clustering, and matching methodology, we observe strong performance with minimal deviation from the optimal IP site selections. In \Cref{fig:DBSCAN_solmaxdata_final}, which depicts the difference for our scalable framework with the maximum data downlink objective, all our solutions met up to $99\%$ of the optimal IP. In \Cref{fig:DBSCAN_solminmaxgap_final} for the minimum maximum-gap objective, we see that all cases result in differences less than 0.08 hours, falling within 95\% of the IP optimum. Examining the majority of these cases, objective values fall within 97\% of the optimum values. These results empirically demonstrate that our scalable pipeline closely matches the results from the globally optimal IP formulation, with only minor trade-offs in strict optimality. This strong performance is not isolated to a small subset but holds across the parameter space.

\begin{figure*}[htbp]
  \centering
  \begin{minipage}[t]{0.4\textwidth}
    \centering
    \includegraphics[scale=0.215]{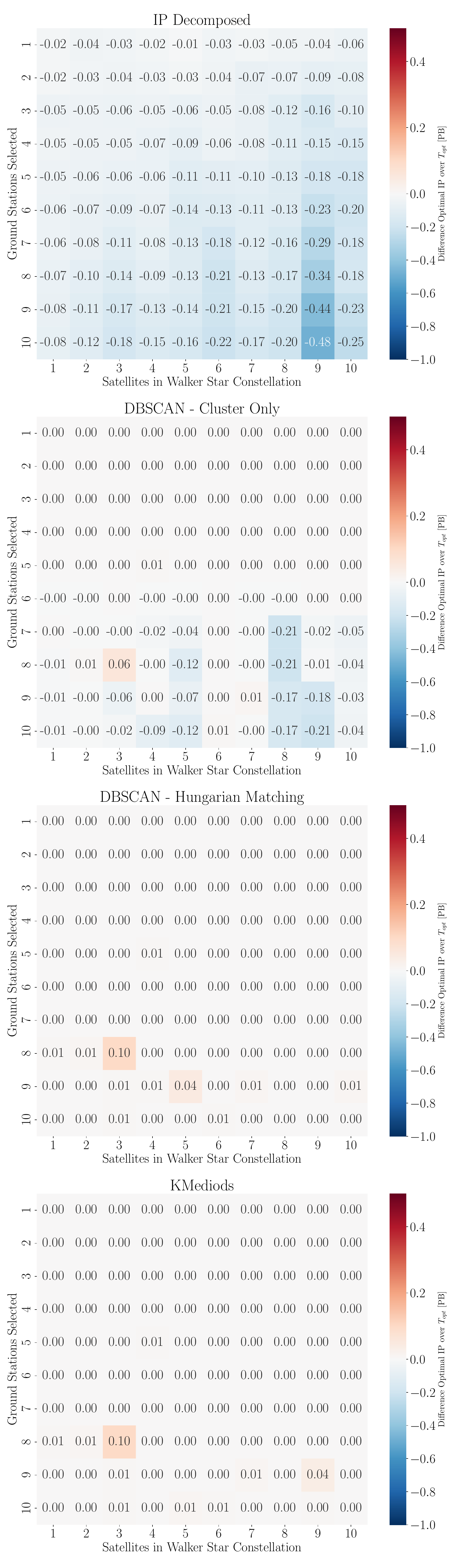}
    \caption{Deviation in performance from optimum in maximum data downlink objective using scalable decomposition applied solely along the time dimension.}
    \label{fig:chunks_false_maxData}
  \end{minipage}%
  \hfill
  \begin{minipage}[t]{0.4\textwidth}
    \centering
    \includegraphics[scale=0.215]{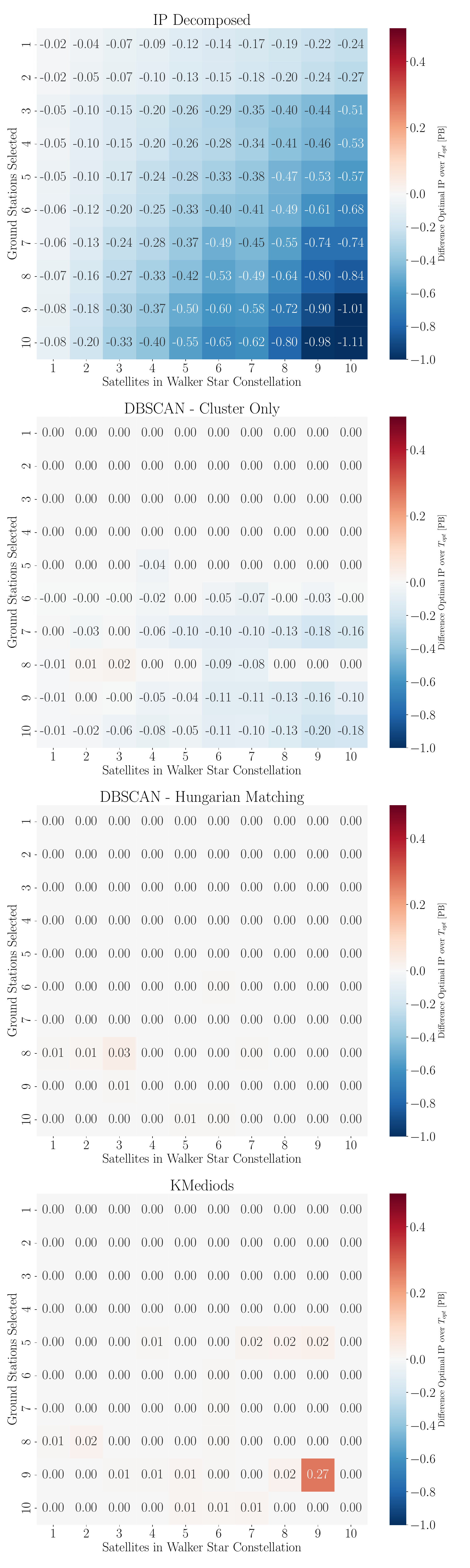}
    \caption{Deviation in performance from optimum in maximum data downlink objective using scalable decomposition applied along both time and satellite dimensions.}
    \label{fig:chunks_true_maxData}
  \end{minipage}
\end{figure*}

\subsection{Stepwise Validation of Scalable Placement Framework}
To rigorously evaluate our scalable GSaaS placement framework, we perform a stepwise validation, comparing intermediate solutions from each stage of the pipeline (decomposition, clustering, and matching) against the ground-truth IP solutions. This allows us to isolate the contribution of each component and understand where deviations from optimality occur. We also test an alternative clustering and matching method, $k$-mediods, in addition to our DBSCAN and Hungarian matching method to test alternative matching mechanisms. Experiments with maximum data downlink objective are shown in Figures \ref{fig:chunks_false_maxData} and \ref{fig:chunks_true_maxData}, while those with minimum maximum-gap are shown in Figures \ref{fig:chunks_False_minmaxgap} and \ref{fig:chunks_true_minmaxgap}.

\subsubsection{Comparability of Methods} It is important to note that not all methods are directly comparable to the ground-truth IP solutions in \Cref{fig:IP_solmaxdata,fig:IP_solminmaxgap}. The decomposed IP only solves smaller subproblems from the full IP formulation, either over shorter simulation interval windows, and when applicable, smaller subsets of the constellation. To compare these with the full IP, we scale the decomposed solutions to the full problem size (e.g., multiplying by the time horizon length and/or number of spacecraft in the constellation). This scaling produces solutions that approximate, but do not strictly satisfy, the original IP constraints. For the minimum maximum-gap objective in \Cref{fig:chunks_False_minmaxgap,fig:chunks_true_minmaxgap}, scaling is unnecessary because the maximum gap time over each time window $T_{sim}^j\in\mathcal{T}$ is assumed to reflect the type of gap times found in both $T_{sim}$ and $T_{opt}$. Still, the reported values represent only the optimal outcomes from each time window $T_{sim}^j$ or each individual satellite $S$, not the global IP. 

For the DBSCAN-only clustering baselines, cluster centers are not restricted to actual ground station locations in $\mathcal{L}^D$. This can lead to the DBSCAN-only solutions occasionally outperforming the true IP by placing “virtual” stations. Once cluster centers are mapped back to GSaaS sites (DBSCAN clustering with Hungarian matching or $k$-Medoids), the solutions become fully comparable to the ground truth. Only the DBSCAN + Hungarian matching and $k$-medoids approaches enforce constraints consistent with the full problem. The full list of methods is outlined in \Cref{tab:methods}.

\begin{table}[htbp]
\centering
\caption{Descriptions of methods evaluated, with comparability to the ground-truth IP solution.}
\vspace{-1em}
\footnotesize
\renewcommand{\arraystretch}{1.2}
\begin{tabular}{@{}p{2.6cm}p{1.1cm}p{3.6cm}@{}}
\toprule
\textbf{Method} & \textbf{Directly compare to IP?} & \textbf{Description}\\
\midrule
IP-Optimal& $-$  & Exact integer programming solution (ground truth) \\
IP-Decomposed & No &IP formulation with temporal/spatial decomposition \\
DBSCAN Only & No & Clustering only (no assignment) \\
DBSCAN + Hungarian & Yes &DBSCAN clustering followed by Hungarian matching \\
$k$-Medoids &Yes & Joint clustering and matching \\
\bottomrule
\end{tabular}
\label{tab:methods}
\vspace{1em}
\end{table}

\subsubsection{Decomposition Levels} 
We first compare our method's performance on different levels of decomposition: temporal decomposition (using shorter time windows only) versus a combined approach of temporal and per-satellite decomposition. Comparing decomposition strategies for the maximum data downlink objective in \Cref{fig:chunks_false_maxData} (temporal only) and \Cref{fig:chunks_true_maxData} (temporal and satellite), we find that using both temporal and satellite-level decomposition yields higher-performing subproblem solutions. By solving at the per-satellite level, the smaller IP formulations can identify locally optimal allocations tailored to each spacecraft, which can then be aggregated into stronger overall results. This effect is visible in the heatmaps of \Cref{fig:chunks_true_maxData}, which show more blue regions\textemdash indicating performance that exceeds the full IP in comparison to the temporal-only case in \Cref{fig:chunks_false_maxData}. We observe similar results for the minimum maximum-gap objective, with higher performing individual solutions in \Cref{fig:chunks_true_minmaxgap} (temporal and satellite) versus \Cref{fig:chunks_False_minmaxgap} (temporal only).

\subsubsection{DBSCAN Clustering}  
\begin{figure*}[htbp]
  \centering
  \begin{minipage}[t]{0.4\textwidth}
    \centering
    \includegraphics[scale=0.215]{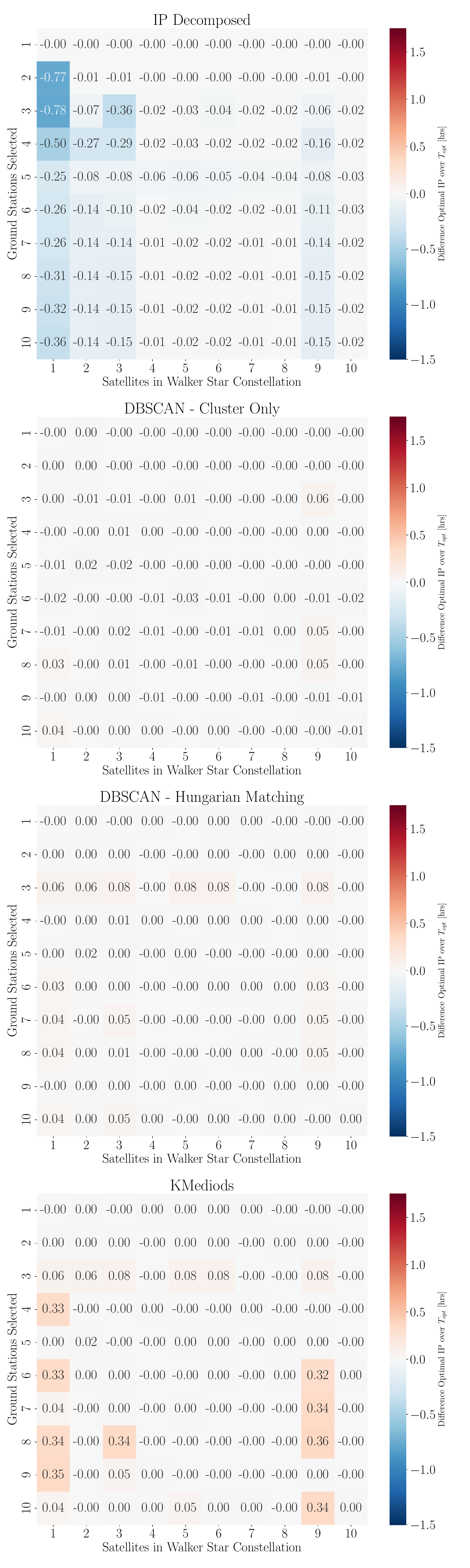}
    \caption{Deviation in performance from optimum in minimum maximum-gap objective using scalable decomposition applied solely along the time dimension.}
    \label{fig:chunks_False_minmaxgap}
  \end{minipage}%
  \hfill
  \begin{minipage}[t]{0.4\textwidth}
    \centering
    \includegraphics[scale=0.215]{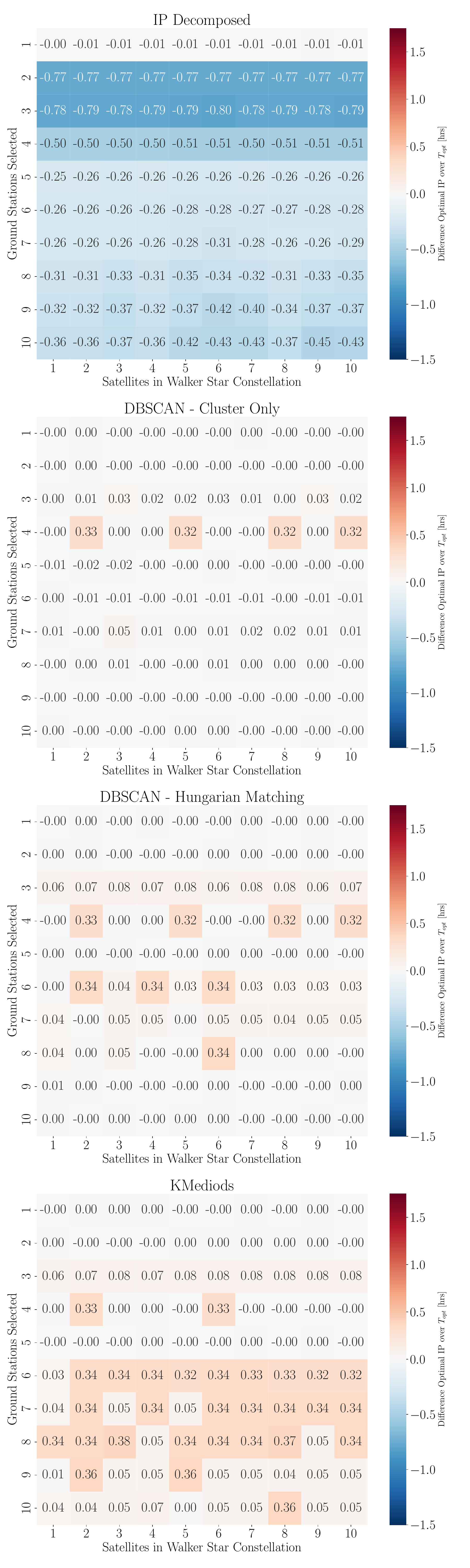}
    \caption{Deviation in performance from optimum in minimum maximum-gap objective using scalable decomposition applied along both time and satellite dimensions.}
    \label{fig:chunks_true_minmaxgap}
  \end{minipage}
\end{figure*}

We next evaluate the effect of clustering decomposed solutions using DBSCAN, prior to mapping clusters to feasible GSaaS sites. At this stage, DBSCAN identifies coverage centroids spanning across all subproblems, unconstrained by real site locations. The quality of DBSCAN-only solutions depends strongly on the decomposition strategy and the objective function. For the maximum data downlink objective, where individual satellite performance dominates over constellation-wide uniformity, temporal+satellite decomposition yields richer and more diverse candidate points, enabling DBSCAN to identify stronger centroids (\Cref{fig:chunks_true_maxData}) compared to temporal-only decomposition (\Cref{fig:chunks_false_maxData}). In this case, solving per-satellite subproblems provides high-quality building blocks that aggregate effectively, while temporal-only decomposition may not yield sufficient diversity in candidate sites, limiting clustering performance when scaled to the full IP. In contrast, for the minimum maximum-gap objective, the trend reverses: temporal-only decomposition (\Cref{fig:chunks_False_minmaxgap}) outperforms temporal+satellite (\Cref{fig:chunks_true_minmaxgap}). Because the gap metric reflects constellation-wide coverage, per-satellite decomposition produces overly specialized solutions that do not recombine as effectively when clustered, resulting in more suboptimal outcomes.  

\subsubsection{Matching to Feasible GSaaS Sites}
At this stage, we enforce the full set of constraints from the original IP formulation by ensuring that clustered solutions are matched to feasible GSaaS sites. We evaluate two approaches: applying Hungarian matching to the DBSCAN cluster centroids, and using $k$-Medoids, which directly restricts centroids to existing points within the subproblem dataset.  Our results show that DBSCAN with Hungarian matching consistently outperforms $k$-Medoids. For the maximum data downlink objective, DBSCAN with Hungarian matching achieves up to $99\%$ of the optimal IP value under both decomposition strategies (temporal-only and temporal-plus-satellite). For the minimum maximum-gap objective, temporal-only decomposition maintains high performance, remaining within $99\%$ of the optimal solution. When combining temporal and satellite decomposition, performance slightly degrades due to clustering over more specialized points; nevertheless, the majority of cases remain within $95\%$ of optimal (e.g., $92\%$ of final scenarios). Performance dips in this setting are driven primarily by the number of selected ground stations rather than constellation size, with underperforming cases concentrated at $n=4,6,8$ in the DBSCAN+Hungarian heatmaps.  In contrast, $k$-Medoids is more restrictive and underperforms across both objectives and decomposition strategies. Because $k$-Medoids requires the initial selection of cluster centers to coincide with GSaaS sites, it limits flexibility in selecting candidate centroids. This limitation becomes especially pronounced as the number of stations increases ($n=6,7,8$ in the minimum maximum-gap heatmaps), where coverage quality degrades more severely compared to DBSCAN with Hungarian matching.

\begin{table*}[tbp]
\centering
\caption{Comparison of ground station selections for the min-max gap objective with $n=6$ satellites. Both clustering/matching methods (DBSCAN+Hungarian and $k$-Medoids) selected identical stations. Maximum gap times: $0.9$ hours (Optimal IP), $1.24$ hours (clustering/matching).}
\label{tab:station_comparison}
\vspace{-1em}
\begin{tabular}{@{}clcclccc@{}}
\toprule
\textbf{Position} & \textbf{Optimal IP} & \textbf{Longitude} & \textbf{Latitude} & \textbf{Clustering Methods\textsuperscript{*}} & \textbf{Longitude} & \textbf{Latitude} & \textbf{Distance (km)} \\
\midrule
1 & Long Beach & $-118.15$ & $33.82$ & Long Beach & $-118.15$ & $33.82$ & $0.0$ \\
2 & Singapore & $103.7$ & $1.32$ & Singapore & $103.7$ & $1.32$ & $0.0$ \\
3 & Svalbard & $15.41$ & $78.23$ & Svalbard & $15.41$ & $78.23$ & $0.0$ \\
4 & Troll & $2.53$ & $-72$ & Troll & $2.53$ & $-72$ & $0.0$ \\
5 & Tolhuin & $-67.12$ & $-54.51$ & Punta Arenas & $-70.87$ & $-52.94$ & $302.9$ \\
6 & Bangalore & $77.37$ & $12.9$ & Mauritius & $57.45$ & $-20.5$ & $4{,}289.9$ \\
\bottomrule
\end{tabular}

\end{table*}

Examining the selected stations of underperforming clustering scenarios of the minimum maximum-gap objective, we found that in most cases, the differences in coverage performance could be traced to a single ground station differing between the optimal and clustering solutions. As an illustrative example, we consider the minimum maximum-gap objective with $n=6$ stations selected from a Walker star constellation of size $\lvert S \rvert = 6$.  \Cref{tab:station_comparison} lists the ground station locations chosen by the optimal IP solution alongside those selected by our clustering and matching methods. In this case, both of our DBSCAN with Hungarian matching and $k$-Medoids methods produced identical station sets. The table highlights both the geographic coordinates and spatial proximity of the clustering-selected stations relative to the optimal solution. While almost all of the same station selections were made as the optimal IP, one single difference in station choice noticeably reduced performance. In this example, the maximum gap time in contact across the constellation is $0.9$ hours for the optimal IP solution, compared to $1.24$ hours for the clustering and matching methods. The fact that such differences stem from a single station suggests that modest, targeted refinements, or individual station swaps for further post-processing could further close performance gaps.

\begin{figure}[htbp]
    \centering
    \includegraphics[trim={0cm 24cm 0cm 0cm},clip,scale=0.3]{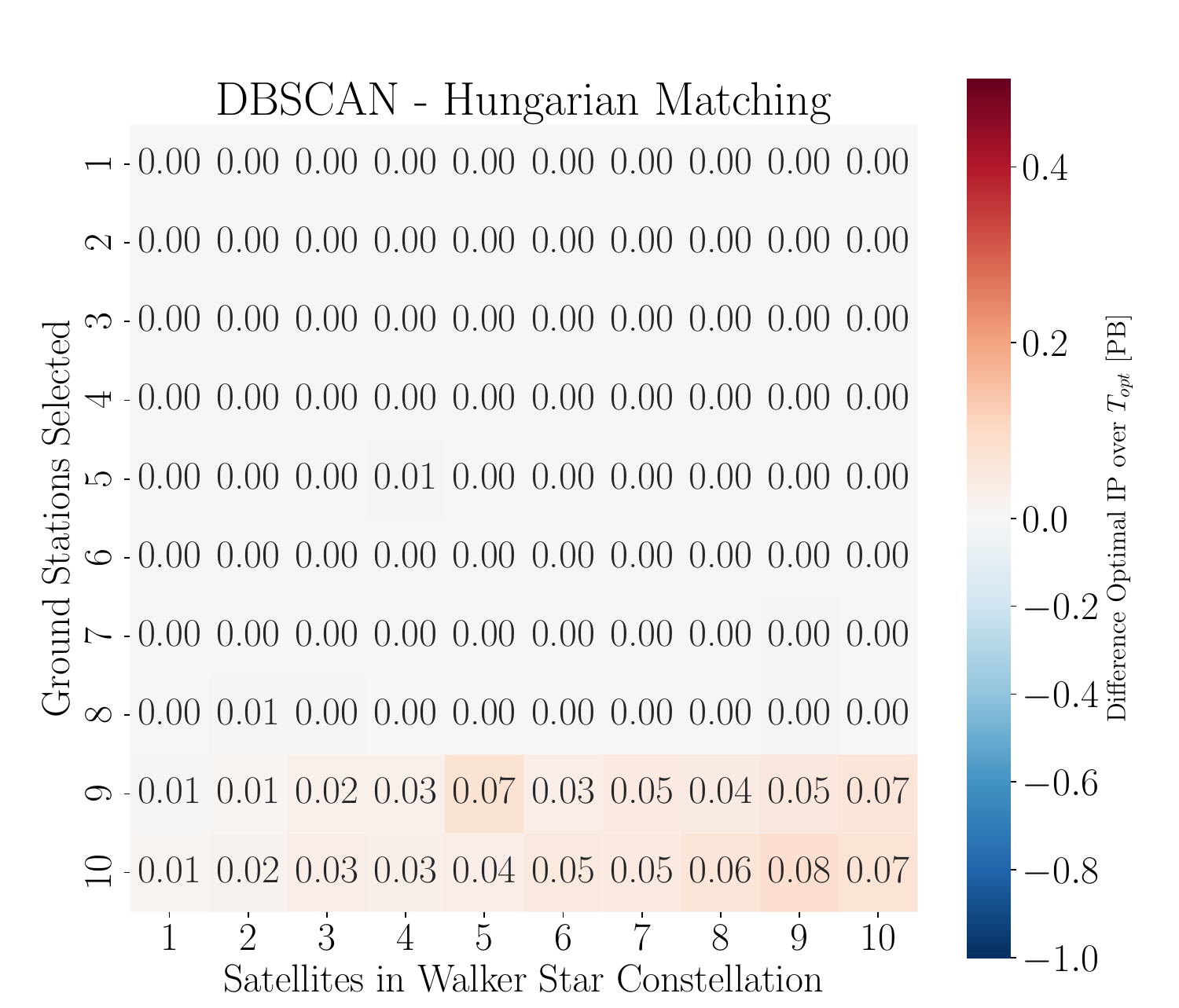}
    \caption{Deviation in solution performance from optimal full-IP solution for the maximum data downlink scenario, when expanding the initial clustering solutions to the full $\mathcal{L}$ ground station candidate site list.}
    \label{fig:all_expansion}
    \vspace{-1em}
\end{figure}

\subsubsection{Expanding Candidate Ground Station Locations}
We now evaluate the final step of our clustering–matching method and how it performs when expanding the candidate ground station set from the design GSaaS subset $\mathcal{L}^D$ (KSAT, Atlas Space) to the full list of candidate sites $\mathcal{L}$ (Atlas Space Operations, AWS Ground Station, KSAT, Leaf Space, and Viasat). For the maximum-data downlink objective, we can benchmark directly against the optimal IP solution, as shown in \Cref{fig:all_expansion}. In this setting, the clustering-based methods nearly match the optimal solutions, demonstrating that if the restricted set $\mathcal{L}^D$ is sufficiently diverse, the clustering and matching framework can extrapolate effectively to select near-optimal ground station locations in the full candidate list $\mathcal{L}$. We observe slight degradation for larger selections, specifically $n=9$ or $10$, but the suboptimality remains modest, still approaching $99\%$ of the original solution. For the min–max gap objective, we are unable to provide direct comparisons in this expanded setting, since the full IP solutions fail to converge reliably.

\subsection{Realistic Scenarios}

We next evaluate our framework on more realistic Earth observation constellations, where the number of satellites, ground station candidates, and overall problem complexity is substantially larger than in the controlled Walker star experiments. In this setting, we test our DBSCAN clustering with Hungarian matching method on three representative operators: Capella Space, ICEYE, and Planet Labs' Flocks, with constellation characteristics summarized in \Cref{tab:EO_Operators}. Once again for the decomposition step, time windows are split into 1-day windows with 12-hour overlaps, and the candidate ground station list $\mathcal{L}^D$ is restricted to the KSAT and Atlas Space GSaaS networks. However, after clustering is performed on these subproblems, we expand back in the matching phase the candidate ground station list to the full list of $\mathcal{L}$, exemplifying the scalability of our approach. We consider both the maximum-data downlink and minimum maximum-gap objectives. Where the full IP solution is computationally tractable, we use it as a benchmark; otherwise, we compare the degradation in solution answers from the initial subproblem decomposition to the final matching step to understand how our method performs at each stage of the scalable decomposition process.

\begin{table}[htbp!]
\centering
\caption{Earth Observation Satellite Constellations.} 
\label{tab:EO_Operators}
\vspace{-1 em}
\begin{tabular}{@{}lccc@{}}
\toprule
Constellation & \# Sats & Altitude (km) & Inclination \\
\midrule
Capella Space & 5 & 525--575 & $45^\circ$, $53^\circ$, $97^\circ$ \\
ICEYE & 34 & 560--580 & $97^\circ$ \\
Planet Flocks & 93 & 500 & $97^\circ$ \\
\bottomrule
\end{tabular}
\end{table}

\begin{table*}[htbp] 
\centering
\caption{Performance of scalable decomposition and clustering methods at each stage in comparison to optimal IP formulation. The approach uses 1-day time windows with per-satellite decomposition, DBSCAN clustering, and Hungarian matching for final GSaaS solutions. Optimizing for maximum data downlink objective, values listed in PB over $T_{opt}$.
Solution $\Delta$ represents the difference between maximum per-satellite decomposed solutions and final Hungarian matching solutions.}
\label{tab:RealData}
\vspace{-1em}
\begin{tabular}{@{}lcccccccccrr@{}}
\toprule
           & \multicolumn{3}{c}{Decomposition} 
           & \multicolumn{3}{c}{DBSCAN Clustering} 
           & \multicolumn{3}{c}{\textbf{Final Hungarian Match}} 
           &  \\
\cmidrule(lr){2-4} \cmidrule(lr){5-7} \cmidrule(lr){8-10}
           Constellation & Min  & Mean & Max
           & Min  & Mean & Max
           & Min  & Mean & Max
           & Optimal & Solution $\Delta$\\
\midrule
Capella Space      &   3.21& 6.10&  6.85&   
4.84& 5.03& 5.06&  4.99& 5.05&  \textbf{5.07}& \textbf{5.23} &1.78 \\
ICEYE      &  12.77&  44.05  &  55.97 & 38.89  &  41.95   &  42.99 & 34.88  &  40.81   &  \textbf{43.03 } & \textbf{43.04} &12.84\\
Flock      &  40.72&  96.34  &  112.23 & 85.80  &  93.86   &  96.33 & 90.32  &  94.54   &  \textbf{96.43 } & \textbf{96.45}&15.8\\
\bottomrule
\end{tabular}
\end{table*}

\begin{table*}[htbp] 
\centering
\caption{Performance of scalable decomposition and clustering methods compared to optimal IP formulation. The approach uses 1-day time windows with per-satellite decomposition, DBSCAN clustering, and Hungarian matching for final GSaaS solutions. All values optimize for minimum maximum-gap between contacts (hours over $T_{opt}$). Solution $\Delta$ represents the difference between minimum per-satellite decomposed solutions and final Hungarian matching solutions. Optimal solutions from IP solvers failed to converge for ICEYE and Planet Labs' FLOCK constellations.}
\label{tab:RealGap}
\vspace{-1em}
\begin{tabular}{@{}lcccccccccrr@{}}
\toprule
           & \multicolumn{3}{c}{Decomposition} 
           & \multicolumn{3}{c}{DBSCAN Clustering} 
           & \multicolumn{3}{c}{\textbf{Final Hungarian Match}} 
           &  \\
\cmidrule(lr){2-4} \cmidrule(lr){5-7} \cmidrule(lr){8-10}
           Constellation & Min  & Mean & Max
           & Min  & Mean & Max
           & Min  & Mean & Max
           & Optimal 
           & Solution $\Delta$
           \\
\midrule
Capella Space      &   0.61& 0.85&  1.14&   
2.33& 2.64& 3.56&  \textbf{2.31}& 2.73&  3.56& \textbf{1.42} & 1.17\\
ICEYE      &  0.53 &  0.88  &  1.31 & 2.43  &  3.31   &  4.50 & \textbf{2.46}  &  3.52   &  4.50 & \textbf{-} & 1.15\\
Flock      &  0.77 &  0.89 &  1.13& 1.44 &  1.55  &  1.58 & \textbf{1.44}  &  1.65   &  1.66 & \textbf{-} &0.31\\
\bottomrule
\end{tabular}
\end{table*}

For the maximum data downlink objective, we were able to directly compare our results to the optimal solutions obtained from the IP solver, allowing us to evaluate how each step of our scalable framework performs with these realistic constellation scenarios. Evaluations of the subproblems within our framework is outlined at three stages: (1) the decomposition results, (2) DBSCAN clustering of the decomposed solutions, and (3) the final Hungarian matching that maps cluster centroids to actual GSaaS station locations. It is important to note that the decomposition and clustering results are not perfectly comparable to the optimal solution, since we normalize decomposition values (by constellation size and time window length), and clustering solutions are generated before enforcing the final station-location constraints. The full numerical results are presented in \Cref{tab:RealData}.

Because performance varies across satellites, decompositions, and clustering parameters, we report ranges (minimum, mean, maximum) for each method. In decomposition, the spread reflects different satellite splits; in clustering, it depends on the choice of neighborhood DBSCAN distance thresholds $\varepsilon$, which ranged from $\{5,10,15,20,25,30,35,40\}$; and the final Hungarian match inherits the variability from clustering. Among these, we are primarily interested in the maximum values, since they represent the “best-case” achievable solutions that should be closest to optimal.

Overall, we find that solution quality decreases when moving from decomposition to clustering. This follows our previous results, since decomposition represents an upper bound (best-case across satellites and time splits), while clustering introduces variability and weaker minimum solutions. Interestingly, the final Hungarian match slightly improves upon the clustering stage in several cases. This suggests that mapping cluster centroids to real GSaaS station locations does not degrade performance as much as expected, and may even benefit from the spatial distribution of actual stations. When compared to the optimal IP solutions, the final matched results are extremely close across most constellations. The main exception is Capella Space, which shows a modest performance gap. This is likely due to its distinct orbital regime with different altitudes and inclination angles for its spacecraft. However, even in this case, the performance of our scalable site selection method remains within a reasonable bound near the optimal IP solution. The approximate value after matching is at 5.07 PB, about 3.06\% lower than the optimal 5.23 PB over $T_{opt}$.

Now, examining the results for the minimum maximum-gap objective in \Cref{tab:RealGap}, we observe trends that differ from the maximum data downlink case. For this objective, the minimum values are the most meaningful, since the goal is to reduce the largest gap between consecutive satellite contacts. The exception is the decomposition-per-satellite metric, where by the objective's definition we select the maximum to represent the maximum gap found in communications across the constellation.

Transitioning from decomposition to DBSCAN clustering introduces a noticeable increase in the maximum gap for both Capella Space and ICEYE constellations. Surprisingly, the largest constellation, FLOCK, exhibits a smaller increase in maximum gap. This is likely due to its satellites having similar orbits, whereas Capella Space and ICEYE have varied orbital altitudes. Capella Space even has differences in inclination angles for its small constellation, explaining why its loss in solution quality is comparable to ICEYE, despite ICEYE having almost eight times more satellites.

Only the Capella Space constellation, with 5 satellites, could be solved using the full IP formulation, highlighting the difficulty of scaling IP solvers for larger constellations and full candidate GSaaS lists for consideration. For Capella Space, comparing the optimal solution (1.42 hours) to the final Hungarian-matched solution (2.31 hours) shows a difference from the true optimum. However, considering the length of the entire optimization window, the approximate solution still provides a good start. For ICEYE and Flock, the full optimal solution is intractable, but the final solutions remain close to the decomposition step considering the size of each constellation. For example, Flock's final solution has a maximum gap of 1.44 hours versus 1.13 hours from the satellite-specific decomposition, indicating only a modest loss in solution quality despite expanding the selected sites to the entire 96 satellite constellation. These results suggest that our decomposition-clustering-matching framework performs especially well for large constellations with similar orbital regimes. Despite ICEYE having eight times more satellites than Capella Space, the solution $\Delta$ remains comparable, indicating this method provides an effective starting point for minimizing maximum gap times in large-scale scenarios.

\section{Conclusions}
This paper addresses the computational challenges of ground station selection for large LEO satellite constellations. We introduce a novel framework that decomposes the GSaaS site selection IP formulation into tractable single-satellite subproblems and leverages clustering and matching techniques to aggregate solutions. This method provides a scalable approach that closely approximates optimal results from the full IP as problem complexity grows. Experiments on both synthetic and real Earth observation constellations demonstrate that the framework produces high-quality ground network designs, achieving near-optimal solutions ($97\%$ or higher) for all maximum data downlink scenarios. For minimizing communication gaps where exact IP methods were intractable for large constellations, such as ICEYE and Planet Labs' Flock satellites, the framework yields minimal degradation, with maximum gap increases being no greater than 0.31 hours for our largest 96-satellite constellation. This approach enables optimization for constellation sizes and candidate station networks far beyond the reach of traditional exact solvers, supporting practical ground network site selection at scale. Our results validate the decomposition, clustering, and matching framework as a powerful tool for next-generation ground station infrastructure, capable of supporting current and emerging mega-constellations without significant loss in network quality. Future work may explore targeted single-station swaps as a local refinement step to further close residual optimality gaps and enhance network performance.

\appendix{}

We provide further background on the external algorithms used in each step of our scalable GSaaS site selection methodology. All of these algorithms are well-established in the literature, and we apply these tools for standard clustering and matching tasks. For reproducibility, source code is available at
\href{https://github.com/gkim65/scalable-ground-station-optimizer}{https://github.com/gkim65/scalable-ground-station-optimizer}.

The \textit{\textbf{DBSCAN}} (Density-Based Spatial Clustering of Applications with Noise) algorithm \cite{DBSCAN} uses point cluster density to identify arbitrarily shaped clusters in spatial datasets. Groupings are focused on points that are closely packed together, while points in low-density regions are treated as noise. The algorithm uses two parameters, a neighborhood radius $\varepsilon$, and a minimum number of points $m_{points}$. Compared to other popular clustering algorithms, the number of clusters do not need to be specified prior to use, and outliers can be effectively detected. For our work, we set $m_{points}$ to be 2 for all cases. The full method is shown in \Cref{alg:DBSCAN}.

\begin{algorithm}[htbp]
\caption{\textit{DBSCAN}. \textit{Base algorithm referenced from the original \textit{DBSCAN} paper~\cite{DBSCAN}} \newline Point dataset $D$, neighborhood radius $\varepsilon$, minimum number of points $m_{points}$.}\label{alg:DBSCAN}
\begin{algorithmic}
\algrenewcommand\algorithmicrequire{\textbf{DBSCAN(D, $\varepsilon$, $m_{points}$)}}
\algrenewcommand\algorithmicensure{\textbf{ExpandCluster}($P$, $\mathcal{N}$, $id$, $\varepsilon$, $m_{points}$):}
\Require 

\State {initialize $id= 0$} \Comment{\textit{Cluster ID}}
\For{each unvisited point $P$ in dataset $D$} 
    \If{$P$ is not yet assigned to cluster}
        \State{$\mathcal{N} \leftarrow$ points within $\varepsilon$ distance of $P$} \Comment{\textit{Neighbors}}
        \If{size($\mathcal{N}$) $< m_{points}$}
            \State{Label $P$ as NOISE}
        \Else
            \State{$id = id +1$}
            \State{\textbf{ExpandCluster}($P$, $\mathcal{N}$, $id$, $\varepsilon$, $m_{points}$)}
        \EndIf
    \EndIf
\EndFor
\newline
\Ensure \Comment{\textit{Run on client/satellite k}}
\State {Assign $P$ to $id$}
\For{ each point $N\in \mathcal{N}$}
\If{$N$ is labeled NOISE}
\State{Assign $N$ to $id$}
\EndIf
\If{$N$ is not yet assigned to any cluster}
\State{Assign $N$ to $id$}
\EndIf
\State{$N_{neighbors} \leftarrow$ points within $\varepsilon$ distance of $N$}
\If{size($N_{neighbors}$) $\geq$ $m_{points}$}
\State{$\mathcal{N}$ =  $\mathcal{N}$ +$N_{neighbors}$} \Comment{\textit{Add all}}
\EndIf
\EndFor

\end{algorithmic}
\end{algorithm}

The \textbf{$k$-Mediods} algorithm \cite{kaufman1990partitioning} is a clustering method similar to $k$-Means, but actual data points (mediods) are selected as final cluster centers. Prior to using the algorithm, the number of clusters $k$ must be defined beforehand. The use of mediods leads the method to be less sensitive to noise and outliers compared to the typical $k$-Means algorithm. The algorithm is prone to differences in final answer based on what random seed is set, as the very first step assigns $k$ random mediods randomly to initialize the algorithm. The algorithm iteratively refines the assignment of data points to medoids and updates the medoids to minimize the total dissimilarity between points and their assigned cluster centers. The full method is shown in \Cref{alg:kmediods}.

\begin{algorithm}[!ht]
\caption{\textit{$k$- Mediods}. \textit{Base algorithm referenced from the original \textit{$k$-Mediods} paper~\cite{kaufman1990partitioning}} \newline Point dataset $D$, number of clusters $k$.}\label{alg:kmediods}
\begin{algorithmic}
\algrenewcommand\algorithmicrequire{\textbf{$k$-Mediods}($D$, $k$)}
\Require 

\State {Select $k$ initial mediods randomly from $D$} 
\State {Assign each non-medoid point $d \in D$ to nearest mediods} 
\State {Initialize best improvement $q_{diff} = 1$} 
\While{$q_{diff} >0$} 
\State{Calculate initial cost $q_o$} \Comment{\textit{Sum of disimmilarities}}
\For{each mediod $m$ and each non-mediod $d \in D$}
    \State{Swap $m$ and $d$}
    \State{Calculate new cost $q_n$} 
    \If{if $q_n<q_o$ and $q_{diff} < q_o-q_n$ }
        \State{$q_{diff} = q_o-q_n$}
        \State{Keep track of swapped $m$ and $d$}
    \EndIf
\EndFor
\If{$q_{diff}>0$}
\State{Perform the best $m$ and $d$ swap found}
\EndIf
\EndWhile

\end{algorithmic}
\end{algorithm}

The \textbf{Hungarian algorithm} \cite{kuhn1955hungarian} is a classical combinatorial optimization technique designed to solve the linear assignment problem efficiently. Given a cost matrix where each entry quantifies the cost of assigning one element from a set to another (for example, workers to jobs), the algorithm computes an assignment that minimizes the total cost, ensuring that each element is uniquely matched.

The method proceeds by first reducing the matrix rows and columns to introduce zero elements, which indicate potential optimal assignments. It then searches for a minimum cover of these zeros with the fewest horizontal and vertical boundaries. If the number of such boundaries equals the matrix dimension, the algorithm determines an optimal assignment directly from these zeros. Otherwise, it adjusts the uncovered elements to generate additional zeros and repeats the process.

This iterative approach guarantees a polynomial-time solution and has become a foundational algorithm in scheduling, resource allocation, and network optimization problems.


\acknowledgements 

This research was supported by the Hertz Foundation and the National Defense Science and Engineering Graduate (NDSEG) Fellowship Program.

\bibliographystyle{IEEEtran}
\bibliography{references.bib}

\begin{thebibliography}{10}
\providecommand{\url}[1]{#1}
\csname url@samestyle\endcsname
\providecommand{\newblock}{\relax}
\providecommand{\bibinfo}[2]{#2}
\providecommand{\BIBentrySTDinterwordspacing}{\spaceskip=0pt\relax}
\providecommand{\BIBentryALTinterwordstretchfactor}{4}
\providecommand{\BIBentryALTinterwordspacing}{\spaceskip=\fontdimen2\font plus
\BIBentryALTinterwordstretchfactor\fontdimen3\font minus \fontdimen4\font\relax}
\providecommand{\BIBforeignlanguage}[2]{{%
\expandafter\ifx\csname l@#1\endcsname\relax
\typeout{** WARNING: IEEEtran.bst: No hyphenation pattern has been}%
\typeout{** loaded for the language `#1'. Using the pattern for}%
\typeout{** the default language instead.}%
\else
\language=\csname l@#1\endcsname
\fi
#2}}
\providecommand{\BIBdecl}{\relax}
\BIBdecl

\bibitem{cleverly2021evolution}
G.~Cleverly, A.~Murray, and B.~Mendler, ``The evolution of ground stations in the new space industry,'' in \emph{ASCEND}, 2021, p. 4042.

\bibitem{carvalho2019optimizing}
R.~Carvalho, ``Optimizing the communication capacity of a ground station network,'' \emph{Journal of Aerospace Technology and Management}, vol.~11, p. e2319, 2019.

\bibitem{vasisht2020distributed}
D.~Vasisht and R.~Chandra, ``A distributed and hybrid ground station network for low earth orbit satellites,'' in \emph{Association for Computing Machinery Workshop on Hot Topics in Networks}, 2020, pp. 190--196.

\bibitem{eddy_optimal_2024}
D.~Eddy, M.~Ho, and M.~J. Kochenderfer, ``Optimal ground station selection for low-earth orbiting satellites,'' in \emph{IEEE Aerospace Conference}, 2025, pp. 1--13.

\bibitem{kopacz2019optimized}
J.~Kopacz, J.~Roney, and R.~Herschitz, ``Optimized ground station placement for a mega constellation using a genetic algorithm,'' in \emph{AIAA/USU Conference on Small Satellites (SSC19-WP1-05)}, 2019.

\bibitem{Kralfallah_Optimizing_2024}
M.~Kralfallah, F.~Wu, A.~Tahir, A.~Oubara, and X.~Sui, ``Optimizing the deployment of ground tracking stations for low earth orbit satellite constellations based on evolutionary algorithms,'' \emph{Remote Sensing}, vol.~16, no.~5, 2024.

\bibitem{owen1998strategic}
S.~H. Owen and M.~S. Daskin, ``Strategic facility location: A review,'' \emph{European Journal of Operational Research}, vol. 111, no.~3, pp. 423--447, 1998.

\bibitem{Azeem_sub_2019}
H.~Azeem, L.~Du, A.~Ullah, M.~A. Mughal, M.~M. Aslam, and M.~Ikram, ``Sub-array based antenna selection scheme for massive {MIMO} in 5g,'' in \emph{International Conference on Cyberspace Data and Intelligence}.\hskip 1em plus 0.5em minus 0.4em\relax Springer, 2019, pp. 38--50.

\bibitem{Wei_multi_2022}
W.~Wei, R.~Yang, H.~Gu, W.~Zhao, C.~Chen, and S.~Wan, ``Multi-objective optimization for resource allocation in vehicular cloud computing networks,'' \emph{IEEE Transactions on Intelligent Transportation Systems}, vol.~23, no.~12, pp. 25\,536--25\,545, 2022.

\bibitem{cardei2006energy}
M.~Cardei and J.~Wu, ``Energy-efficient coverage problems in wireless ad-hoc sensor networks,'' \emph{Computer Communications}, vol.~29, no.~4, pp. 413--420, 2006.

\bibitem{tlili2017swarm}
T.~Tlili, M.~Harzi, and S.~Krichen, ``Swarm-based approach for solving the ambulance routing problem,'' \emph{Procedia Computer Science}, vol. 112, pp. 350--357, 2017.

\bibitem{hu_joint_2018}
M.~Hu, W.~Liu, K.~Peng, X.~Ma, W.~Cheng, J.~Liu, and B.~Li, ``Joint routing and scheduling for vehicle-assisted multi-drone surveillance,'' \emph{IEEE Internet of Things Journal}, vol.~PP, pp. 1--1, 10 2018.

\bibitem{franti2025designing}
P.~Fr{\"a}nti, S.~Sieranoja, and T.~Laatikainen, ``Designing a clustering algorithm for optimizing health station locations,'' \emph{International Journal of Health Geographics}, vol.~24, no.~1, p.~4, 2025.

\bibitem{imai2023analysis}
R.~A.~M. Imai, C.~B. da~Cunha, and C.~S. Guazzelli, ``An analysis of the impact of demand aggregation on the solution quality for facility location problems,'' \emph{Transportes}, vol.~31, no.~3, pp. e2801--e2801, 2023.

\bibitem{buijs2025effective}
R.~Buijs, R.~van~der Mei, E.~Dugundji, and S.~Bhulai, ``An effective aggregation heuristic for capacitated facility location problems with many demand points,'' \emph{Computers and Operations Research}, p. 107153, 2025.

\bibitem{DBSCAN}
A.~Ram, S.~Jalal, A.~Jalal, and K.~Manoj, ``A density based algorithm for discovering density varied clusters in large spatial databases,'' \emph{International Journal of Computer Applications}, vol.~3, 06 2010.

\bibitem{kaufman1990partitioning}
L.~Kaufman, ``Partitioning around medoids (program pam),'' \emph{Wiley Series in Probability and Statistics}, vol. 344, pp. 68--125, 1990.

\bibitem{gurobi}
{Gurobi Optimization, LLC}, ``{Gurobi Optimizer Reference Manual},'' \url{https://www.gurobi.com}, 2024.

\bibitem{lougee2003common}
R.~Lougee-Heimer, ``The common optimization interface for operations research: promoting open-source software in the operations research community,'' \emph{IBM Journal of Research and Development}, vol.~47, no.~1, pp. 57--66, 2003.

\bibitem{scikit-learn}
F.~Pedregosa, G.~Varoquaux, A.~Gramfort, V.~Michel, B.~Thirion, O.~Grisel, M.~Blondel, P.~Prettenhofer, R.~Weiss, V.~Dubourg, J.~Vanderplas, A.~Passos, D.~Cournapeau, M.~Brucher, M.~Perrot, and E.~Duchesnay, ``Scikit-learn: Machine learning in {P}ython,'' \emph{Journal of Machine Learning Research}, vol.~12, pp. 2825--2830, 2011.

\bibitem{schubert2022fast}
E.~Schubert and L.~Lenssen, ``Fast k-medoids clustering in rust and python,'' \emph{Journal of Open Source Software}, vol.~7, no.~75, p. 4183, 2022.

\bibitem{kuhn1955hungarian}
H.~W. Kuhn, ``The hungarian method for the assignment problem,'' \emph{Naval Research Logistics Quarterly}, vol.~2, no. 1-2, pp. 83--97, 1955.

\bibitem{2020SciPy-NMeth}
P.~Virtanen, R.~Gommers, T.~E. Oliphant, M.~Haberland, T.~Reddy, D.~Cournapeau, E.~Burovski, P.~Peterson, W.~Weckesser, J.~Bright, S.~J. {van der Walt}, M.~Brett, J.~Wilson, K.~J. Millman, N.~Mayorov, A.~R.~J. Nelson, E.~Jones, R.~Kern, E.~Larson, C.~J. Carey, {\.I}.~Polat, Y.~Feng, E.~W. Moore, J.~{VanderPlas}, D.~Laxalde, J.~Perktold, R.~Cimrman, I.~Henriksen, E.~A. Quintero, C.~R. Harris, A.~M. Archibald, A.~H. Ribeiro, F.~Pedregosa, P.~{van Mulbregt}, and {SciPy 1.0 Contributors}, ``{{SciPy} 1.0: Fundamental Algorithms for Scientific Computing in Python},'' \emph{Nature Methods}, vol.~17, pp. 261--272, 2020.

\bibitem{vallado2006revisiting}
D.~Vallado, P.~Crawford, R.~Hujsak, and T.~Kelso, ``Revisiting {S}pacetrack {R}eport \#3,'' in \emph{AIAA/AAS {Astrodynamics} Specialist Conference and Exhibit}, 2006.

\end{thebibliography}

\newpage
\thebiography
\begin{biographywithpic}{Grace Ra Kim}{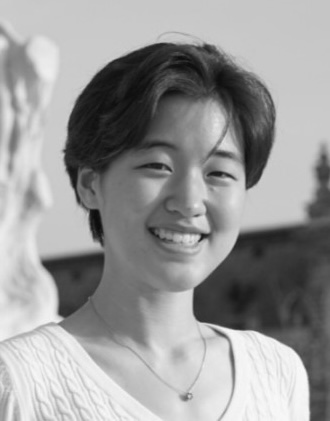}
is a PhD student in the Stanford Intelligent Systems Laboratory (SISL) in the department of Aeronautics and Astronautics at Stanford University, as a Hertz and NDSEG Fellow. She received her M.Phil. in Advanced Computer Science at the University of Cambridge as a Marshall Scholar in 2024, and graduated cum laude with high honors in a B.S. in Engineering Sciences from Harvard University in 2023. Her research interests lie at the intersection of space traffic management, safe autonomous systems in space, and space policy.

\end{biographywithpic}

\begin{biographywithpic}{Duncan Eddy}{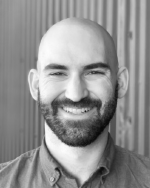}
is a postdoctoral research fellow in the Stanford Intelligent Systems Laboratory and the Executive Director of the Center for AI Safety at Stanford University. His research focuses on decision-making in safety-critical, climate, and space systems. He received B.S. in Mechanical Engineer from Rice University in 2013, and PhD in Aerospace Engineering from Stanford University in 2021. Prior to returning to Stanford he was the Director of Space Operations at Capella Space Corporation where he built a fully automated constellation tasking and delivery system. He later founded and led the Constellation Management and Space Safety Organization at Amazon's Project Kuiper.
\end{biographywithpic} 

\begin{biographywithpic}{Vedant Srinivas}{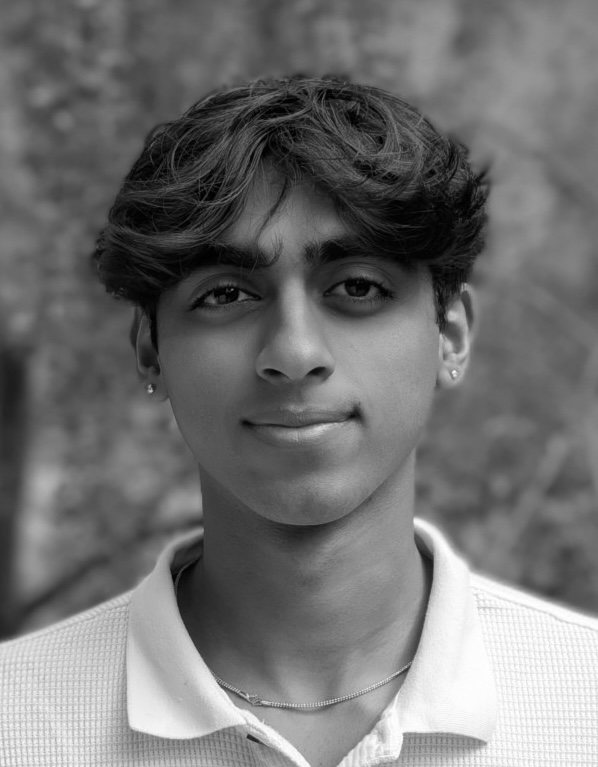}
is an undergraduate in the Stanford Intelligent Systems Laboratory in the department of Computer Science at Stanford University. His research interests include optimization and decision-making for autonomous systems, with experience as an AI Software Engineer Intern at Salesforce AI Research. He is also the co-founder of IyarkAI, a startup deploying real-time wildlife monitoring systems with the Washington Department of Transportation, and previously worked at the UC Davis Road Ecology Center on edge-AI systems for wildlife detection and collision prevention.
\end{biographywithpic}

\begin{biographywithpic}{Mykel J. Kochenderfer}{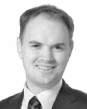}
 is an Associate Professor of Aeronautics and Astronautics and Associate Professor, by courtesy, of Computer Science at Stanford University. He is the director of the Stanford Intelligent Systems Laboratory, conducting research on advanced algorithms and analytical methods for the design of robust decision making systems. Prior to joining the faculty in 2013, he was at MIT Lincoln Laboratory where he worked on airspace modeling and aircraft collision avoidance. He received his Ph.D. from the University of Edinburgh in 2006 where he studied at the Institute of Perception, Action and Behaviour in the School of Informatics. He received B.S. and M.S. degrees in computer science from Stanford University in 2003.

\end{biographywithpic}

\end{document}